\documentclass[preprint,showpacs,prc,aps,<floats,superscriptaddress]{revtex4-1}
\usepackage{epsfig}
\usepackage{graphicx}
\usepackage{longtable}
\usepackage{amsmath}
\usepackage{amssymb}
\newcommand{\agev}    {\mbox{$A$~GeV}}               
\newcommand{\gevc}    {\mbox{GeV$/c$}}

\newcommand{\mevc}    {\mbox{MeV$/c$}}

\newcommand{\rb}[1]   {\mbox{\textrm{\scriptsize #1}}}

\newcommand{\lam}     {\ensuremath{\Lambda}}

\newcommand{\pimin}   {\ensuremath{\pi^-}}
\newcommand{\piplus}  {\ensuremath{\pi^+}}
\newcommand{\kmin}    {\ensuremath{\textrm{K}^-}}
\newcommand{\kplus}   {\ensuremath{\textrm{K}^+}}

\newcommand{\pt}      {\ensuremath{p_{\rb{T}}}}
\newcommand{\plab}    {\ensuremath{p_{\rb{lab}}}}
\newcommand{\mt}      {\ensuremath{m_{\rb{T}}}}

\newcommand{\dedx}    {\ensuremath{\textrm{d}E/\textrm{d}x}}
\newcommand{\dndy}    {\ensuremath{\textrm{d}N/\textrm{d}y}}

\newcommand{\nwound}  {\ensuremath{\langle N_{\rb{w}} \rangle}}

\begin{document}

\title{System-size and centrality dependence of charged kaon and pion production
in nucleus-nucleus
collisions at 40\agev~and 158\agev~beam energy}


 \affiliation{NIKHEF, Amsterdam, Netherlands.}
 \affiliation{Department of Physics, University of Athens, Athens,
 Greece.}
 \affiliation{E\"otv\"os Lor\'ant University, Budapest, Hungary. }
 \affiliation{Wigner Research Centre for Physics, Hungarian Academy of Sciences,
 Budapest, Hungary.}
 \affiliation{MIT, Cambridge, USA.}
 \affiliation{H.~Niewodnicza\'nski Institute of Nuclear Physics, Polish Academy
of Sciences, Cracow,
 Poland.}
 \affiliation{GSI Helmholtzzentrum f\"{u}r Schwerionenforschung, Darmstadt,
 Germany.}
 \affiliation{Joint Institute for Nuclear Research, Dubna,
 Russia.}
 \affiliation{Fachbereich Physik der Universit\"{a}t, Frankfurt,
 Germany.}
 \affiliation{CERN, Geneva, Switzerland.}
 \affiliation{Institute of Physics, Jan Kochanowski University, Kielce,
 Poland.}
 \affiliation{Fachbereich Physik der Universit\"{a}t, Marburg,
 Germany.}
 \affiliation{Max-Planck-Institut f\"{u}r Physik, Munich,
 Germany.}
 \affiliation{Inst. of Particle and Nuclear Physics, Charles Univ., Prague,
Czech
 Republic.}
 \affiliation{Nuclear Physics Laboratory, University of Washington, Seattle, WA,
 USA.}
 \affiliation{Atomic Physics Department, Sofia Univ. St. Kliment Ohridski,
Sofia,
 Bulgaria.}
 \affiliation{Institute for Nuclear Research and Nuclear Energy, BAS, Sofia,
 Bulgaria.}
 \affiliation{Department of Chemistry, Stony Brook Univ. (SUNYSB), Stony Brook,
 USA.}
 \affiliation{Institute for Nuclear Studies, Warsaw, Poland.}
 \affiliation{Institute for Experimental Physics, University of Warsaw, Warsaw,
 Poland.}
 \affiliation{Faculty of Physics, Warsaw University of Technology, Warsaw,
 Poland.}
 \affiliation{Rudjer Boskovic Institute, Zagreb, Croatia.}

 \author{T.~Anticic}
        \affiliation{Rudjer Boskovic Institute, Zagreb, Croatia.}
 \author{B.~Baatar}
       \affiliation{Joint Institute for Nuclear Research, Dubna,
 Russia.}
 \author{D.~Barna}
       \affiliation{Wigner Research Centre for Physics, Hungarian Academy of Sciences,
 Budapest, Hungary.}
 \author{J.~Bartke}
        \affiliation{H.~Niewodnicza\'nski Institute of Nuclear Physics, Polish
Academy of Sciences, Cracow,
 Poland.}
 \author{H.~Beck}
      \affiliation{Fachbereich Physik der Universit\"{a}t, Frankfurt,
 Germany.}
 \author{L.~Betev}
       \affiliation{CERN, Geneva, Switzerland.}
 \author{H.~Bia{\l}\-kowska}
       \affiliation{Institute for Nuclear Studies, Warsaw, Poland.}
 \author{C.~Blume}
      \affiliation{Fachbereich Physik der Universit\"{a}t, Frankfurt,
 Germany.}
 \author{M.~Bogusz}
         \affiliation{Faculty of Physics, Warsaw University of Technology,
Warsaw,
 Poland.}
 \author{B.~Boimska}
      \affiliation{Institute for Nuclear Studies, Warsaw, Poland.}
 \author{J.~Book}
      \affiliation{Fachbereich Physik der Universit\"{a}t, Frankfurt,
 Germany.}
 \author{M.~Botje}
        \affiliation{NIKHEF, Amsterdam, Netherlands.}
 \author{P.~Bun\v{c}i\'{c}}
       \affiliation{CERN, Geneva, Switzerland.}
 \author{T.~Cetner}
       \affiliation{Faculty of Physics, Warsaw University of Technology, Warsaw,
 Poland.}
 \author{P.~Christakoglou}
 \affiliation{NIKHEF, Amsterdam, Netherlands.}
 \author{P.~Chung}
      \affiliation{Department of Chemistry, Stony Brook Univ. (SUNYSB), Stony
Brook,
 USA.}
 \author{O.~Chvala}
      \affiliation{Inst. of Particle and Nuclear Physics, Charles Univ., Prague,
Czech
 Republic.}
 \author{J.G.~Cramer}
      \affiliation{Department of Chemistry, Stony Brook Univ. (SUNYSB), Stony
Brook,
 USA.}
\author{P.~Dinkelaker}
     \affiliation{Fachbereich Physik der Universit\"{a}t, Frankfurt,
 Germany.}
 \author{V.~Eckardt}
       \affiliation{Max-Planck-Institut f\"{u}r Physik, Munich,
 Germany.}
 \author{Z.~Fodor}
       \affiliation{Wigner Research Centre for Physics, Hungarian Academy of Sciences,
 Budapest, Hungary.}
 \author{P.~Foka}
       \affiliation{GSI Helmholtzzentrum f\"{u}r Schwerionenforschung, Darmstadt,
 Germany.}
 \author{V.~Friese}
       \affiliation{GSI Helmholtzzentrum f\"{u}r Schwerionenforschung, Darmstadt,
 Germany.}
 \author{M.~Ga\'zdzicki}
       \affiliation{Fachbereich Physik der Universit\"{a}t, Frankfurt,
 Germany.}
       \affiliation{Institute of Physics, Jan Kochanowski University,
Kielce,
 Poland.}
 \author{K.~Grebieszkow}
        \affiliation{Faculty of Physics, Warsaw University of Technology,
Warsaw,
 Poland.}
 \author{C.~H\"{o}hne}
       \affiliation{GSI Helmholtzzentrum f\"{u}r Schwerionenforschung, Darmstadt,
 Germany.}
 \author{K.~Kadija}
       \affiliation{Rudjer Boskovic Institute, Zagreb, Croatia.}
 \author{A.~Karev}
       \affiliation{CERN, Geneva, Switzerland.}
\author{M.~Kliemant}
     \affiliation{Fachbereich Physik der Universit\"{a}t, Frankfurt,
 Germany.}
 \author{V.I.~Kolesnikov}
      \affiliation{Joint Institute for Nuclear Research, Dubna,
 Russia.}
 \author{T.~Kollegger}
     \affiliation{Fachbereich Physik der Universit\"{a}t, Frankfurt,
 Germany.}
 \author{M.~Kowalski}
       \affiliation{H.~Niewodnicza\'nski Institute of Nuclear Physics, Polish
Academy of Sciences, Cracow,
 Poland.}
 \author{D.~Kresan}
        \affiliation{GSI Helmholtzzentrum f\"{u}r Schwerionenforschung, Darmstadt,
 Germany.}
 \author{A.~Laszlo}
        \affiliation{Wigner Research Centre for Physics, Hungarian Academy of Sciences,
 Budapest, Hungary.}
 \author{R.~Lacey}
        \affiliation{Department of Chemistry, Stony Brook Univ. (SUNYSB), Stony
Brook,
 USA.}
 \author{M.~van~Leeuwen}
        \affiliation{NIKHEF, Amsterdam, Netherlands.}
\author{B.~Lungwitz}
     \affiliation{Fachbereich Physik der Universit\"{a}t, Frankfurt,
 Germany.}
 \author{M.~Mackowiak}
       \affiliation{Faculty of Physics, Warsaw University of Technology, Warsaw,
 Poland.}
 \author{M.~Makariev}
      \affiliation{Institute for Nuclear Research and Nuclear Energy, BAS,
Sofia,
 Bulgaria.}
 \author{A.I.~Malakhov}
       \affiliation{Joint Institute for Nuclear Research, Dubna,
 Russia.}
 \author{M.~Mateev}
      \affiliation{Atomic Physics Department, Sofia Univ. St. Kliment Ohridski,
Sofia,
 Bulgaria.}
 \author{G.L.~Melkumov}
      \affiliation{Joint Institute for Nuclear Research, Dubna,
 Russia.}
 \author{M.~Mitrovski}
        \affiliation{Fachbereich Physik der Universit\"{a}t, Frankfurt,
 Germany.}
 \author{St.~Mr\'owczy\'nski}
        \affiliation{Institute of Physics, Jan Kochanowski University, Kielce,
 Poland.}
 \author{V.~Nicolic}
       \affiliation{Rudjer Boskovic Institute, Zagreb, Croatia.}
 \author{G.~P\'{a}lla}
         \affiliation{Wigner Research Centre for Physics, Hungarian Academy of Sciences,
 Budapest, Hungary.}
 \author{A.D.~Panagiotou}
         \affiliation{Department of Physics, University of Athens, Athens,
 Greece.}
 \author{W.~Peryt}
        \affiliation{Faculty of Physics, Warsaw University of Technology,
Warsaw,
 Poland.}
 \author{J.~Pluta}
        \affiliation{Faculty of Physics, Warsaw University of Technology,
Warsaw,
 Poland.}
 \author{D.~Prindle}
        \affiliation{Nuclear Physics Laboratory, University of Washington,
Seattle, WA,
 USA.}
 \author{F.~P\"{u}hlhofer}
       \affiliation{Fachbereich Physik der Universit\"{a}t, Marburg,
 Germany.}
 \author{R.~Renfordt}
       \affiliation{Fachbereich Physik der Universit\"{a}t, Frankfurt,
 Germany.}
 \author{C.~Roland}
       \affiliation{MIT, Cambridge, USA.}
 \author{G.~Roland}
        \affiliation{MIT, Cambridge, USA.}
 \author{M.~Rybczy\'nski}
        \affiliation{Institute of Physics, Jan Kochanowski University, Kielce,
 Poland.}
 \author{A.~Rybicki}
        \affiliation{H.~Niewodnicza\'nski Institute of Nuclear Physics, Polish
Academy of Sciences, Cracow,
 Poland.}
 \author{A.~Sandoval}
       \affiliation{GSI Helmholtzzentrum f\"{u}r Schwerionenforschung, Darmstadt,
 Germany.}
 \author{N.~Schmitz}
        \affiliation{Max-Planck-Institut f\"{u}r Physik, Munich,
 Germany.}
 \author{T.~Schuster}
         \affiliation{Fachbereich Physik der Universit\"{a}t, Frankfurt,
 Germany.}
 \author{P.~Seyboth}
         \affiliation{Max-Planck-Institut f\"{u}r Physik, Munich,
 Germany.}
 \author{F.~Sikl\'{e}r}
         \affiliation{Wigner Research Centre for Physics, Hungarian Academy of Sciences,
 Budapest, Hungary.}
 \author{E.~Skrzypczak}
         \affiliation{Institute for Experimental Physics, University of Warsaw,
Warsaw,
 Poland.}
 \author{M.~Slodkowski}
        \affiliation{Faculty of Physics, Warsaw University of Technology,
Warsaw,
 Poland.}
 \author{G.~Stefanek}
        \affiliation{Institute of Physics, Jan Kochanowski University, Kielce,
 Poland.}
 \author{R.~Stock}
       \affiliation{Fachbereich Physik der Universit\"{a}t, Frankfurt,
 Germany.}
 \author{H.~Str\"{o}bele}
       \affiliation{Fachbereich Physik der Universit\"{a}t, Frankfurt,
 Germany.}
 \author{T.~Susa}
        \affiliation{Rudjer Boskovic Institute, Zagreb, Croatia.}
 \author{M.~Szuba}
       \affiliation{Faculty of Physics, Warsaw University of Technology, Warsaw,
 Poland.}
 \author{M.~Utvi\'{c}}
     \affiliation{Fachbereich Physik der Universit\"{a}t, Frankfurt,
 Germany.}
 \author{D.~Varga}
         \affiliation{E\"otv\"os Lor\'ant University, Budapest, Hungary. }
 \author{M.~Vassiliou}
          \affiliation{Department of Physics, University of Athens, Athens,
 Greece.}
 \author{G.I.~Veres}
        \affiliation{Wigner Research Centre for Physics, Hungarian Academy of Sciences,
 Budapest, Hungary.}
 \author{G.~Vesztergombi}
        \affiliation{Wigner Research Centre for Physics, Hungarian Academy of Sciences,
 Budapest, Hungary.}
 \author{D.~Vrani\'{c}}
        \affiliation{GSI Helmholtzzentrum f\"{u}r Schwerionenforschung, Darmstadt,
 Germany.}
 \author{Z.~W{\l}odarczyk}
       \affiliation{Institute of Physics, Jan Kochanowski University, Kielce,
 Poland.}
 \author{A.~Wojtaszek-Szwarc}
       \affiliation{Institute of Physics, Jan Kochanowski University, Kielce,
 Poland.}

\collaboration{NA49 Collaboration} \noaffiliation

\date{\today}

\begin{abstract}
Measurements of charged pion and kaon production are presented 
in centrality selected Pb+Pb collisions at 40\agev~and 158\agev~beam energy
as well as in semi-central C+C and Si+Si interactions at 
40\agev. Transverse mass spectra, rapidity spectra and
total yields are determined as a function of centrality.
The system-size and centrality dependence of relative strangeness 
production in nucleus-nucleus collisions at 40\agev~and 158\agev~beam energy
are derived from the data presented here and published data for C+C 
and Si+Si collisions at 158\agev~beam energy.
At both energies a steep increase with centrality
is observed for small systems followed by a weak rise or even saturation
for higher centralities. This behavior is compared to calculations using 
transport models (UrQMD and HSD), a percolation model and the core-corona approach.
\end{abstract}


\maketitle

\section{Introduction}

Strangeness production has always been an important observable for probing the
state of matter created in heavy-ion collisions. The yields of the various 
strange hadrons,
properly normalized to those in elementary nucleon-nucleon interactions, are
expected to exhibit different behavior in 
partonic than in hadronic environments \cite{rafelski-mueller}.  
The number of created strange quark-antiquark pairs per participating nucleon is
significantly higher in central Pb+Pb collisions than in p+p interactions at all
collision energies. A maximum in the yield of the  \kplus~relative to that of 
the \piplus~in central Pb+Pb collisions was observed by NA49 at
about 30\agev~beam energy \cite{na49_energy,na49_20-30} and interpreted as 
evidence for the onset of deconfinement. Microscopic models such as HSD and 
UrQMD do not predict a maximum. In the statistical model this structure can 
only be reproduced by including additional features like those introduced in
\cite{ABS}, i.e. high mass resonances beyond the ones listed in \cite{PDG}
and a significant contribution by $\sigma$ meson states.
There are no indications for a similar
non-monotonic energy dependence of this ratio in p+p interactions. In order to
better understand the observed maximum in the energy dependence observed in
central Pb+Pb collisions, measurements of the system
size dependence of strangeness production are needed for different
strangeness carriers and at various energies. 

NA49 has published data on the system-size dependence of hyperon
($\Lambda$ and $\Xi$) \cite{na49_multis} and (anti-)proton yields
and distributions \cite{Milica} in Pb+Pb collisions 
at 40\agev~and 158\agev\ beam energy. Results on pions, kaons, $\phi$ mesons and
\lam\ hyperons in C+C and Si+Si at 
158\agev\ were reported in Ref. \cite{na49_size}. 
Pion and kaon production was also studied in p+p interactions at
158 \gevc~\cite{Fischer_pi,Fischer_K}. The present study completes these results
by a measurement of the centrality dependence of pion and kaon production in
Pb+Pb collisions at 40\agev~and 158\agev~beam energy as well as for C+C (66\%
most central) and Si+Si (29\% most central) interactions at 40\agev. These
results provide a link between the centrality selected data on pion and kaon
production at lower (SIS, AGS) and higher (RHIC) energy.
At SIS and AGS energies kaon and pion production was studied in
centrality selected Au+Au collisions as well as in collisions of
smaller systems such as Si+Al \cite{sis,ags,Wang}. Here, a rather
linear increase is observed  for relative 
kaon production  with the number of participants. On the other hand, at RHIC
energies a clear saturation behavior is found for collisions
with more than approximately 60 nucleons participating in the collision
\cite{phenix,star}. The same observation has also been made at top SPS energies
comparing central C+C, Si+Si and Pb+Pb collisions \cite{na49_size}.

Statistical models based on the grand-canonical ensemble (G-CE) have been
used to describe strange particle yields in central
collisions of heavy nuclei at high energies \cite{BGKMS,ABS,BMG}. 
For smaller reaction volumes, because of the lower number of produced strange 
quarks, restrictions due to strangeness conservation arise, which
suppress strangeness production with respect to the infinite volume limit. 
This is taken into account by using canonical ensembles
in the statistical model calculations.
The transition from the canonical to the grand-canonical ensemble would thus 
result in a characteristic volume dependence of strangeness production
\cite{rafelski-danos,redlich}.
When comparing statistical model results to experimental data in terms of
reaction volume, the latter has to be related to an experimental observable. In
Refs. \cite{redlich,Cle98} the reaction volume was assumed to be proportional to the
number of wounded nucleons (see section III). However with this approach the data
at SPS energies and above could not be described satisfactorily \cite{percolation}.  
Good agreement can be
achieved, if the relevant volume is subject to additional geometric constraints
which restrict strangeness conservation to subvolumes, as realized for example in the
percolation models. The corresponding calculations suggest that not only for small
systems several subvolumes are formed, but that also in central collisions not
all nucleon-nucleon collisions are included in the central volume
\cite{percolation}. A simpler implementation of this ansatz is realized in the
so-called core-corona model \cite{core-corona,Bozek}. It is built on the assumption
that the fireball created in nuclear collisions is composed of a core, which has
the same properties as a very central collision system, and a
corona, which is a superposition of independent nucleon-nucleon interactions. 
The sizes of core and corona can be defined by the
condition that the corona is formed by those nucleon-nucleon collisions in which
both partners interact only once during the whole collision process. The
fraction of single scatterings is calculated using straight line geometry as
described in the Glauber model \cite{glauber}. This approach was recently 
successfully applied to the system-size dependence of strangeness production at
RHIC energies \cite{becattini,werner-rhic} and for the above mentioned hyperon
production at SPS \cite{na49_multis}. The common features of such model 
calculations at fixed energy are a fast increase of relative strangeness
production with system size for small reaction volumes (below approximately 60 
participating nucleons) and eventual saturation for large system sizes.

\section{The NA49 experiment}
The NA49 detector is a large acceptance hadron spectrometer at the CERN
SPS~\cite{NA49NIM}. The main components 
are four large time projection chambers (TPCs) and two super-conducting dipole
magnets with a one meter 
vertical gap, aligned in a row, and a total bending power of 9~Tm. 
The magnetic field was set a factor of four lower for
data taking at 40\agev\ than at 158\agev.
Two two-meter long TPCs (VTPCs) inside the magnets each with 72 pad-rows along
the beam direction allow for precise tracking, momentum determination, 
vertex reconstruction, and particle identification (PID) by the measurement of
the energy loss (\dedx) in the detector gas with a resolution of 6\%. 
The other two TPCs (MTPCs) downstream of the
magnets have large dimensions (4~m x 4~m x 1.2~m, 90 pad-rows) and provide 
additional momentum resolution for high momentum particles as well as PID by
\dedx~measurement with a resolution of around 4\%. A
momentum resolution in the range $\sigma(p)/p^{2} = 
(0.3-7) \cdot~10^{-4}~(\gevc)^{-1}$ is achieved.
Two time-of-flight scintillator arrays of 891 pixels each (TOF),
situated just behind the MTPCs symmetrically on either side
of the beam axis, add additional K/$\pi$ separation power in the 
laboratory momentum range from one to ten \gevc~(near mid-rapidity 
for kaons). 
A veto-calorimeter (VCAL) placed 20 meters downstream of the target accepts
all beam particles and projectile fragments as well as most of the spectator
neutrons and protons \cite{NA49NIM}. The geometrical acceptance of VCAL was
readjusted for each beam energy by means of a collimator for optimum projectile
spectator coverage. VCAL is used  for off-line (on-line) selection of event 
centrality in Pb+Pb (C+C and Si+Si) collisions. The NA49 detector is described
in detail in reference~\cite{NA49NIM}.

For Pb+Pb collisions a primary SPS beam of 40\agev\ and 158\agev\
was directed onto a Pb target while for C+C and Si+Si interactions a
fragmented Pb beam of 40\agev\ was used. The
fragments were identified by magnetic rigidity ($Z/A= 0.5$) and by a
pulse-height measurement in a scintillation counter in the beam. For C+C and Si+Si
interactions beam particles with $Z$ around 6 and $Z$ around 14, respectively, 
were selected online. A refined $Z$ selection was done offline, which yielded a pure 
C-beam, whereas for the "Si-beam"
the different charges could not be separated well enough, thus a mixture of ions
with $Z$ = 13, 14, and 15 (intensity ratio 34:44:22) were accepted. The target
disks had thicknesses of  10~mm (carbon), 5~mm (silicon) and 200~$\mu$m (lead)
corresponding to 7.9\%, 4.4\%, and 0.5\% interaction lengths for the respective 
beam particles.

The Pb (light ion) beam passed through a quartz Cherenkov (scintillator)
detector, from which the start signal for the time-of-flight measurement was
obtained, and three stations of multi-wire proportional chambers which measured
the trajectory and energy loss of individual beam particles. A minimum bias
trigger for Pb+Pb interactions was derived from the signal of a gas Cherenkov
device right behind the target. Only interactions which reduce the beam charge
and thus the signal seen by this detector by at least 10\% were accepted. The
interaction cross section thus defined is 5.7 b at both energies. The trigger
for the C+C and Si+Si data taking at 40\agev~was based on the energy deposited
in VCAL. The 66\% most central C+C and 29\% most central Si+Si interactions were
selected by requiring an energy in VCAL below the imposed threshold. These fractions 
are determined using the inelastic A+A cross sections (10\% uncertainty), the 
trigger cross sections (5\% uncertainty), and a simulation of the VCAL response.

\section{Data analysis}

The data sample used for this analysis was collected in specific minimum bias runs 
of the NA49 experiment and thus are different from the central data presented 
earlier [2]. 
The recorded minimum bias Pb+Pb collisions were divided into five
consecutive centrality bins: C0 - C4 (see Tab. \ref{centrality}
and Ref. \cite{ALaszlo}). The centrality 
selection is based on the forward going energy of projectile spectators as
measured in VCAL. 
We quantify centrality by the fraction of cross-section according to intervals
of forward going spectator energy. For each centrality interval a 
characteristic quantity, the mean value of "wounded
nucleons" \nwound~is calculated. A nucleon is considered wounded, if its 
interaction occurs
in the nuclear overlap volume. \nwound~is determined by generating
VCAL spectra with events from the 
VENUS 4.12 Monte Carlo code \cite{venus}. The simulation took into account the 
energy resolution of the calorimeter and contributions of participants to the
energy recorded by VCAL. After cross calibration of
experimental and simulated spectra for effects of the experimental trigger in
the most peripheral centrality bin the meqan number of interacting (wounded) nucleons
(\nwound) was extracted from the model data for each of the selected cross section 
fractions \cite{Laszlo_method}. The latter were selected to be
identical at 40\agev~and 158\agev. Thus only insignificant differences of 
\nwound~are observed between both energies 
except for the 10\% discrepancy in the most peripheral centrality interval which we 
attribute to the slightly different online trigger
conditions. Table \ref{centrality} summarizes these numbers together
with the number of analyzed events. For convenience we quote here the 
centrality of the C+C and Si+Si collisions at 158 \agev~\cite{na49_size}
which are used for comparison in this paper:
the most central 15.3\% ($\nwound = 14 \pm$2) and 12.2\% ($\nwound = 37 \pm$3), 
respectively.

\begin{table}[t]
 \begin{ruledtabular}
  \begin{tabular}{ r r r r r r}
    $E_{beam}$ & Class & Centrality & $\nwound$ & 
    $N_{event} / 10^3$ \\
    (GeV) & & (\%) & & \\
    \hline
    40$A$ & C+C & 0-66 & 8.8 $\pm$1.1 & 135 \\
    \hline
    40$A$ & Si+Si & 0-29 & 30.5 $\pm$3.5 & 65 \\
\hline\hline
40$A$ & C0 & 0-5 &356 $\pm$1& 13\\
40$A$  & C1 & 5-12.5 &292 $\pm$2&  23\\
40$A$  & C2 & 12.5-23.5 &212 $\pm$3&  34\\
40$A$  & C3 & 23.5-33.5&144 $\pm$4& 33\\
40$A$  & C4 & 33.5-43.5& 93 $\pm$7&  32\\
\hline\hline
158$A$ & C0 & 0-5 &357 $\pm$1 &  15\\
158$A$ & C1 & 5-12.5 &288 $\pm$2&  24\\
158$A$ & C2 & 12.5-23.5 &211 $\pm$3&  37\\
158$A$ & C3 & 23.5-33.5&146 $\pm$4& 35\\
158$A$ & C4 & 33.5-43.5& 85 $\pm$7&  35\\
\hline
  \end{tabular}
 \end{ruledtabular}
\vspace{-0.2cm}
 \caption
 {\label{centrality} Overview of the analyzed data and centrality classes for
minimum bias Pb+Pb collisions.
 The centrality is given as fraction of the total inelastic cross
 section. $\nwound$ is the average number of wounded
 nucleons per event, and $N_{event}$ the number of analyzed events.}
\end{table}

The recorded Pb+Pb events are contaminated by background due to non-target 
interactions. This contamination was minimized by cuts on the fitted
vertex position and the quality of the fit. It is negligible for near central
collisions and amounts to less than 5\% for the most peripheral collisions
(see~\cite{ALaszlo}). Similar cuts were applied to the C+C and Si+Si collision
events. Details for the latter are given in Table 4.1 of \cite{Benjamin_diploma}.
 
The track finding efficiency and \dedx\ resolution were optimized by track
quality criteria, some of which are different for pions and kaons. These 
criteria are meant to select phase space regions with low track densities,
where the background and efﬁciency corrections are small and approximately uniform.
Only those particles were considered for which the transverse momentum kick 
due to the magnetic field points into 
the same direction as the azimuthal angle projected onto the deflection plane.
In addition the number of reconstructed points on the particle track had 
to exceed 50\% of all measurable points. Further track cuts on azimuthal angles 
with respect to the bending plane and for a minimum of measured and/or measurable 
points are meant to obtain high quality track samples. These track cuts were
chosen differently 
for different particle species and data samples to optimize phase space 
coverage: For negatively charged hadrons and kaons azimuthal angles within 
$\pm$30 degrees with respect to the bending plane were selected except for 
kaons in the C+C and Si+Si events where $\pm$45 degrees were allowed to enhance 
statistics. Tracks of negatively charged hadrons in C+C and Si+Si events were 
required to have at least 10 potential points in the TPCs, while in Pb+Pb 
events a minimum of 30 measured points was asked for. Kaon tracks in C+C 
and Si+Si events had to have a minimum of 50 measured points in all TPCs 
and at least 1 potential point in the VTPCs. In Pb+Pb events they were 
required to have a minimum of 50 measured points in the MTPCs and at least 
1 measured point in the VTPCs, if the  number of measurable points was larger 
than 10. Finally kaon candidate tracks in the sample of Pb+Pb events
were required to point to the main interaction 
vertex with a precision of 2 cm in the $x$- (bending) and 1 cm in the 
$y$-direction.


While kaon spectra were determined by means of particle identification, pion spectra
were obtained by subtracting from all negatively charged particles the contributions 
from electrons, kaons, and anti-protons (see below). Kaons were identified using the 
energy loss (\dedx) information from the TPCs and time-of-flight measurements
in the phase space covered by the TOF walls. 
For the analysis based on \dedx~information alone, raw yields of K$^+$ and 
K$^-$ were extracted from fits to the \dedx~distributions in bins of laboratory 
(\plab) momentum (11 bins from 5 to 50 \gevc~in logarithmic scale for the 158$A$ GeV 
data set and 10 bins from 4 to 32 \gevc~in logarithmic scale 
for the 40$A$ GeV data set), and of transverse momentum (\pt) in 20 bins 
from 0 to 2 GeV/$c$ in linear scale. See \cite{p-ap_paper} for a detailed 
description of the fitting method. The energy loss information from the MTPCs 
(VTPCs and MTPCs) was used in the analysis of minimum bias Pb+Pb (C+C and Si+Si) 
collisions. Bins with less than 500 particles were not used in the further analysis 
to ensure sufficient statistics for the fits to the \dedx~distributions. This 
cut limits the accessible range in \pt~ to values below 1 to 1.5 \gevc~depending
 on \plab. The phase space in \plab~and thus the acceptance in rapidity $y$ is 
limited by the fact that only for momenta from \plab~=~4~GeV/$c$ to \plab~=~50~GeV/$c$  
the mean energy loss of kaons is sufficiently separated from those of pions and 
protons to allow for the extraction of their yield by unfolding of the \dedx~
distributions. This restriction limits the acceptance of kaons to forward 
rapidities. The resulting distributions were corrected for geometrical acceptance, 
in-flight decay of kaons, and reconstruction efficiency. The first two corrections 
were determined by detector simulations using GEANT \cite{geant}. The reconstruction 
efficiency is track density dependent only for Pb+Pb collisions. It was shown 
earlier \cite{na49_size} that the corrections are negligible for C+C and Si+Si 
interactions. The correction factor was determined by embedding simulated tracks 
into real events in three centrality bins ranging from (0-12.5)~\%, (12.5-33.5)~\%, 
and $> 33.5$~\%. Track losses are of order (5-10)~\% for central collisions 
at 158$A$ GeV and drop to only (1-2)~\% in the most peripheral collisions 
and for 40$A$~GeV beam energy. Systematic errors were determined by varying 
track selection criteria and the parameters in the \dedx~unfolding procedure. 
In particular we decreased the minimum number of measured points from 50 to 30 
and used an azimuthal cut of $\pm$50 instead of $\pm$30 degrees. The asymmetry 
parameter in the \dedx~unfolding procedure, which accounts for the asymmetry 
of the distributions of the truncated mean (for details see \cite{p-ap_paper}), 
was varied within the limits 0.6 and 0.8. Different combinations of these 
parameter settings resulted in the systematic errors indicated in Tables II and III. 
The spectra at mid-rapidity were obtained using the combined \dedx~and TOF 
information (Pb+Pb only; statistics were not sufficient for C+C and Si+Si).
The distributions of $m^2$ calculated from the recorded time of flight 
($\sigma_{\rm TOF} \sim 65$~ps) were parametrized in bins of laboratory momentum as 
the sum of three Gaussians  for pions, kaons and protons plus exponential tails 
stemming from unresolved double hits or coincidence with $\gamma$ conversion in the 
scinitillators. For laboratory momenta below 2.5~\gevc, this measurement is 
sufficient to separate kaons from 
pions and protons with high efficiency ($\sim 95$\%) and high purity ($> 96$\%). 
For higher momenta, the independent measurements of time of flight and specific 
energy loss are used simultaneously. 
In the two-dimensional plane of $m^2$ and \dedx, kaons were identified by an elliptical 
region centred at the expectation values and with half-axes equal to $2.5~\sigma_{m^2}$ 
and $2.5~\sigma_{\rm \dedx}$, respectively. 
In addition, regions where the local kaon contribution, according to the parametrisations 
of $m^2$ and \dedx, is below 0.7 were excluded.  
Each track in the kaon identification area entered the transverse monentum spectrum with 
a specific weight accounting for the efficiency of particle identification and for the 
efficiency of the time-of-flight detector. 
The identification efficiency is defined by the ratio of the integral of the 
two-dimensional kaon distribution in $m^2$ and \dedx~inside the identification area 
to its total integral. It does not depend on the centrality class and is determined 
as function of track momentum. For the kaon identification employed here, the efficiency 
is about 95\% and varies only slightly with momentum. The contamination by pions and 
protons is below 1\% for all momenta and is not corrected for.
Efficiency losses of the time-of-flight detectors are caused by multiple hits in one 
scintillator and quality cuts on the deposited charge. They were determined from the 
data themselves for each channel and in each centrality class separately. The average 
efficiency is 82\% and varies slightly  with centrality. Finally, the transverse mass 
spectra were corrected for the geometrical acceptance and the in-flight kaon decay, 
which was obtained by a detector simulation with GEANT.

The separation of negatively charged pions from kaons and anti-protons in the 
acceptance of the NA49 detectors by means of the energy loss measurement is 
limited to high transverse momenta (\pt~larger than 500 MeV/$c$ at mid-rapidity 
and larger than 200 MeV/$c$ at $y_{cm}$=1 for 158\agev~beam energy). Therefore a 
different method was chosen for the evaluation of $\pi^-$ production 
(see \cite{na49_20-30}). Yields of all negatively charged hadrons (h$^-$ 
from the primary vertex were extracted 
in bins of rapidity and \pt~assuming the pion mass. The contamination by K$^-$, 
$\bar{\rm{p}},~$e$^-$, and by particles from the decay of strange particles and 
from secondary interactions, both reconstructed at the primary vertex, was estimated 
by simulation and subtracted. The calculation of theses corrections is based on 
events generated by the VENUS model \cite{venus}, propagated through the NA49 detector 
using GEANT, and processed by the standard reconstruction chain. The ratio 
of reconstructed to all simulated particles yields correction factors in all bins 
of rapidity and \pt~which account for all background tracks in the negatively 
charged hadron yields. Care was taken to account for differences between simulated 
and measured hadron yields by appropriate scaling factors. These were obtained 
from a comparison of the simulated yields with measured data. Where measured 
 yields are not available, particle ratios from statistical model calculations 
were used to infer the corresponding yields. In general distributions of background 
particle from VENUS events were scaled by a global factor. The 
largest corrections to the pion yields arise from K$^-$ in Pb+Pb collisions
(158$A$ GeV) at high \pt~(30\% for \pt$~>$ 1.2 GeV/$c$) where pion yields 
are low and kaons prevail. The K$^-$ distributions were therefore scaled 
differentially (and iteratively) 
in rapidity and \pt~in order to reproduce the shapes of the 
measured spectra. The differences in the correction factors on the negatively 
charged particles from K$^-$ was found to be in the range of 1-3\%. As this 
is a very small change, we found it justified to use global factors for the 
other contributions. At mid-rapidity and low \pt~($<$ 0.1 \gevc) corrections 
amount to 30\% - 40\% due to contributions from (secondary) electrons 
(158$A$ GeV). This large correction requires a careful evaluation as the 
contribution relies fully on the GEANT calculation. We therefore identified 
and removed electrons from the track samples in both the simulation and in 
the data (for the latter based on \dedx~measurements). Results agree 
within 5\% - 10\% in the corresponding bins of rapidity and transverse 
momentum and within 2\% - 3\% for the overall pion yield. For the overall 
evaluation of  systematic errors on the pion yield we performed further 
analyses with different track selection parameters (minimum of 50 instead 
of 30 measured points), reversed magnetic field, and a 
different data set with slightly varied settings for the online selection 
of minimum bias events. Variations of the pion yield are less than 5\% in 
centrality bins C0-C3 and 10\% in bin C4. This higher error in the latter 
bin is due to the uncertainty of the event selection for the most peripheral
bin. In C+C and Si+Si collisions at 40\agev~\pimin~were analyzed with both 
methods (h$^-$ and \dedx). The differences were found to be below 5\%.

The subtraction method is not well suited for the determination of the
$\pi^{+}$ yields at our energies, because positively charged particles have 
a large contribution of protons. However, for Pb+Pb collisions yields at 
mid-rapidity and full phase space can be inferred using the feed-down corrected 
$\pi^{+}/\pi^{-}$ ratio from the combined \dedx\ and TOF analysis in an 
acceptance which is a banana shaped region in the rapidity-\pt~plane within 
$0 < \pt~< 1400$ \mevc~and $2.1 < y < 3.1~(0 < \pt~< 1300$ \mevc~and
$2.5 < y < 5.0$) at 40\agev~(158\agev).
This method relies on the assumption that the \pt~ and rapidity 
distributions of $\pi^{-}$ and $\pi^{+}$ have the same shapes. In VENUS 
events this assumption leads to differences of 2\% between the direct and 
scaled $\pi^{+}$ multiplicities. 
The extracted $\pi^{+}/\pi^{-}$ ratios are $0.93\pm~0.2$ and $0.90\pm~0.1$ 
at 158\agev~and 40\agev~beam energies, respectively and are subject to a 
4\% systematic uncertainty, which is derived from the difference between 
two different methods (see \cite{na49_energy} and \cite{na49_20-30}). We
assume that this ratio is centrality independent. For C+C and
Si+Si collisions the $\pi^{-}/\pi^{+}$ ratios were determined
at 158\agev~beam energy (0.99 and 1.02) in \cite{na49_size}.
We use these same values at 40\agev. More
details on the analysis of minimum bias Pb+Pb collisions can be
found in \cite{peter_phd}. The C+C and Si+Si analyses of pion and kaon yields at
40\agev~are described in \cite{Benjamin_diploma} and \cite{Michael},
respectively.

\section{Results}

\subsection{Transverse mass spectra}

\begin{figure*}[h]
 \includegraphics[width=0.22\textwidth,height=0.32\textwidth]
 {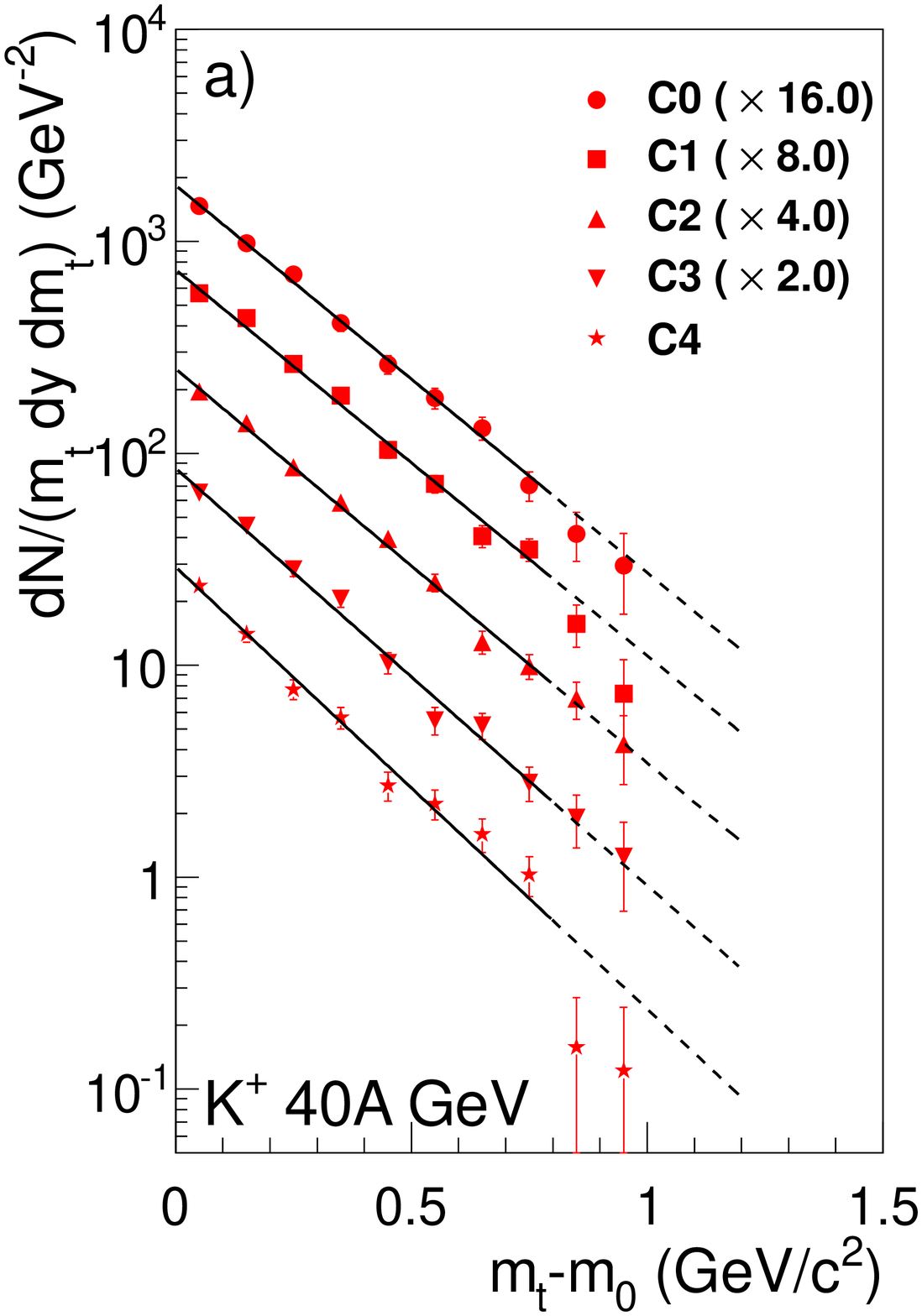}
 \hfill
\vspace{-0.3cm}
 \includegraphics[width=0.22\textwidth]{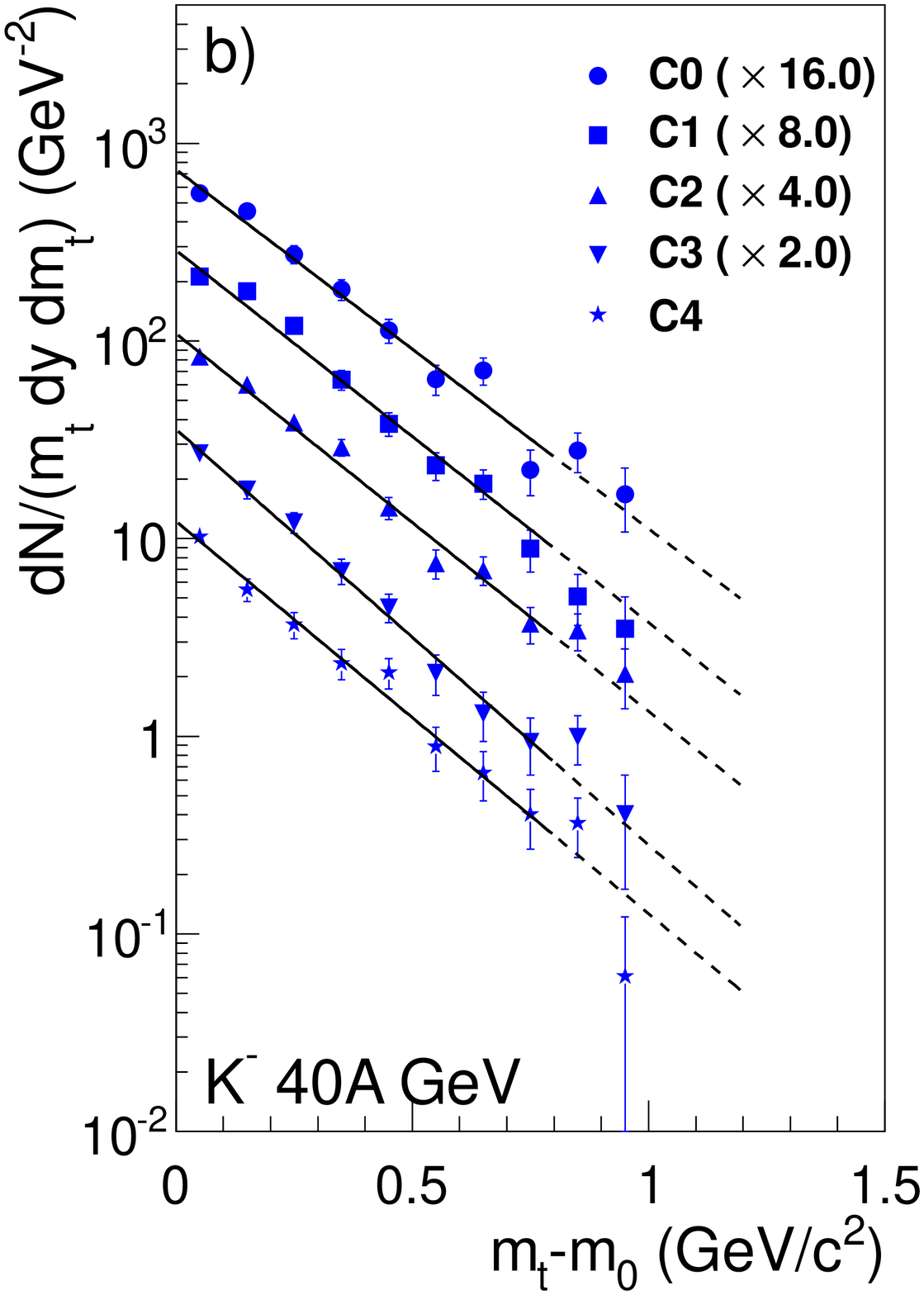}
 \hfill
 \includegraphics[width=0.22\textwidth]{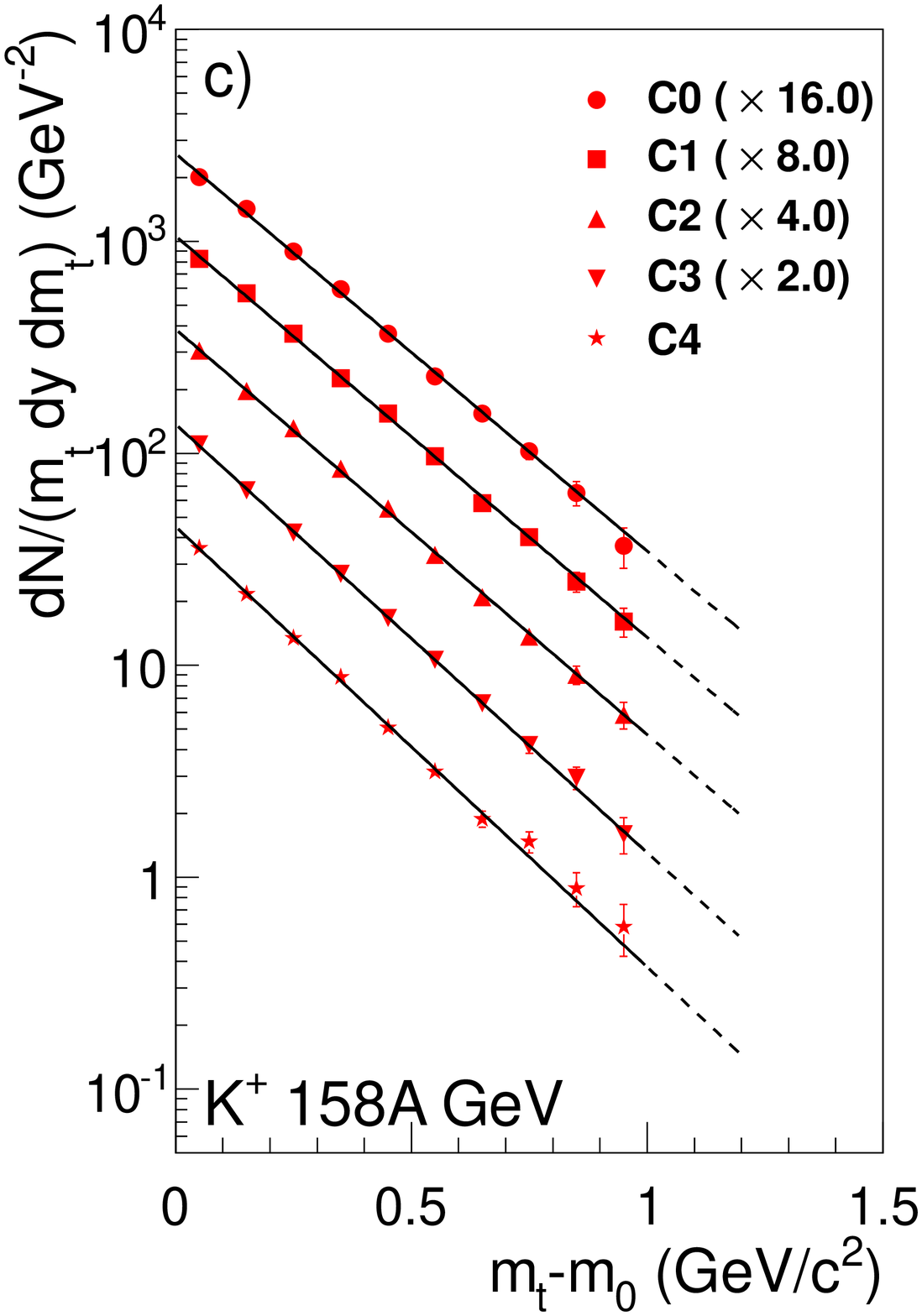}
 \hfill
 \includegraphics[width=0.22\textwidth]{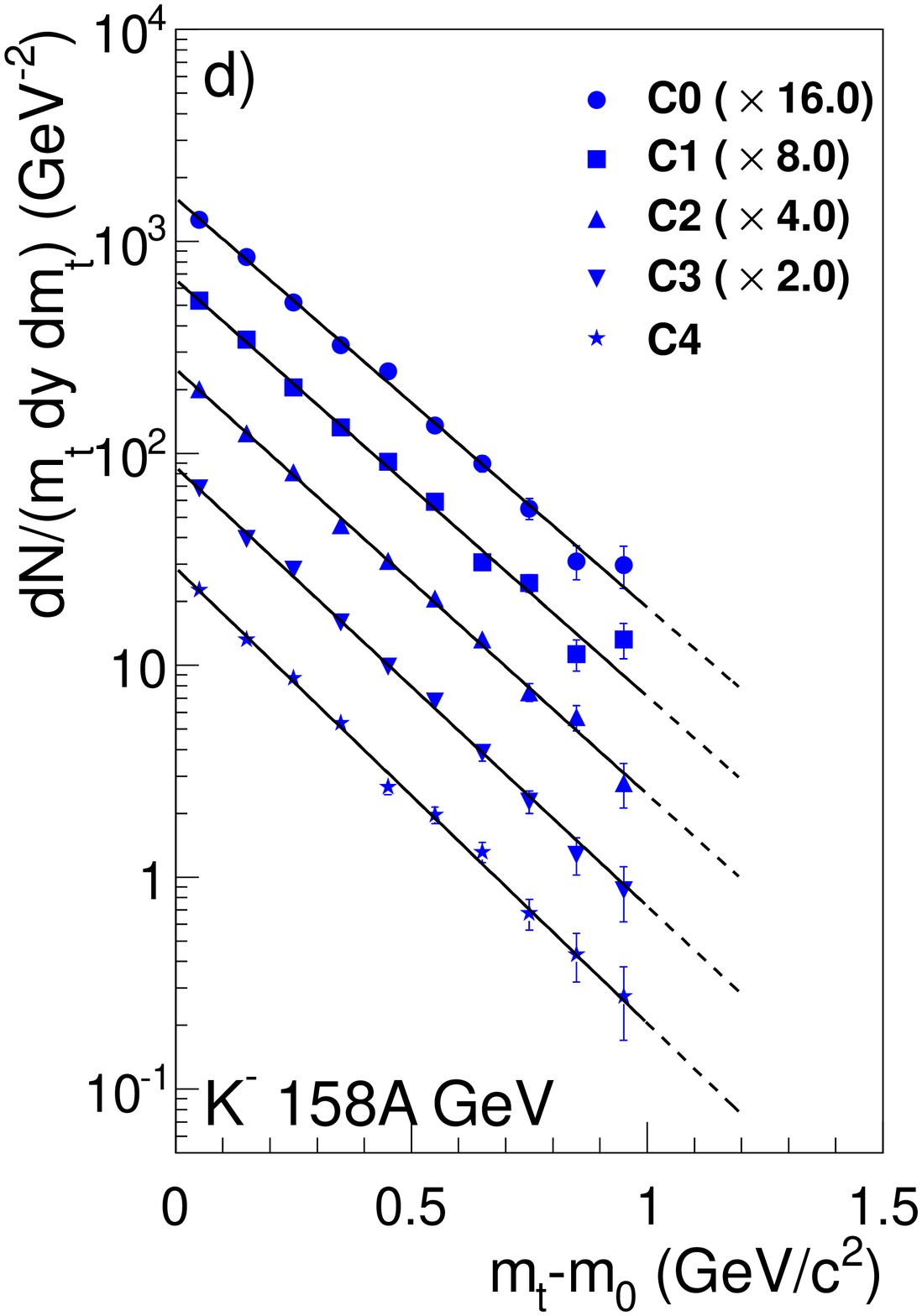}
 \vspace{-0.2cm}
\caption{\label{pt-spectra-K}(Color online) Transverse mass spectra at
mid-rapidity from the combined \dedx\ and TOF analysis for K$^{+}$
and K$^{-}$ in Pb+Pb collisions at 40\agev~($-0.1 < y < 0.1$)
and 158\agev~($-0.2
< y < 0.2$) beam energy. Lines show exponential fits for
0.0~GeV/$c^{2}$ $< \mt-m_{0} <$ 0.8~GeV/$c^{2}$ (40\agev) and
0.0~GeV/$c^{2}$ $< \mt-m_{0} <$ 1.0~GeV/$c^{2}$ (158\agev).
Different centralities are scaled for clarity. The error bars indicate the statistical
uncertainty.}
\end{figure*}

\begin{figure}[h]
\begin {center}
\begin{minipage}[t]{10cm}
 \includegraphics[width=0.44\textwidth]{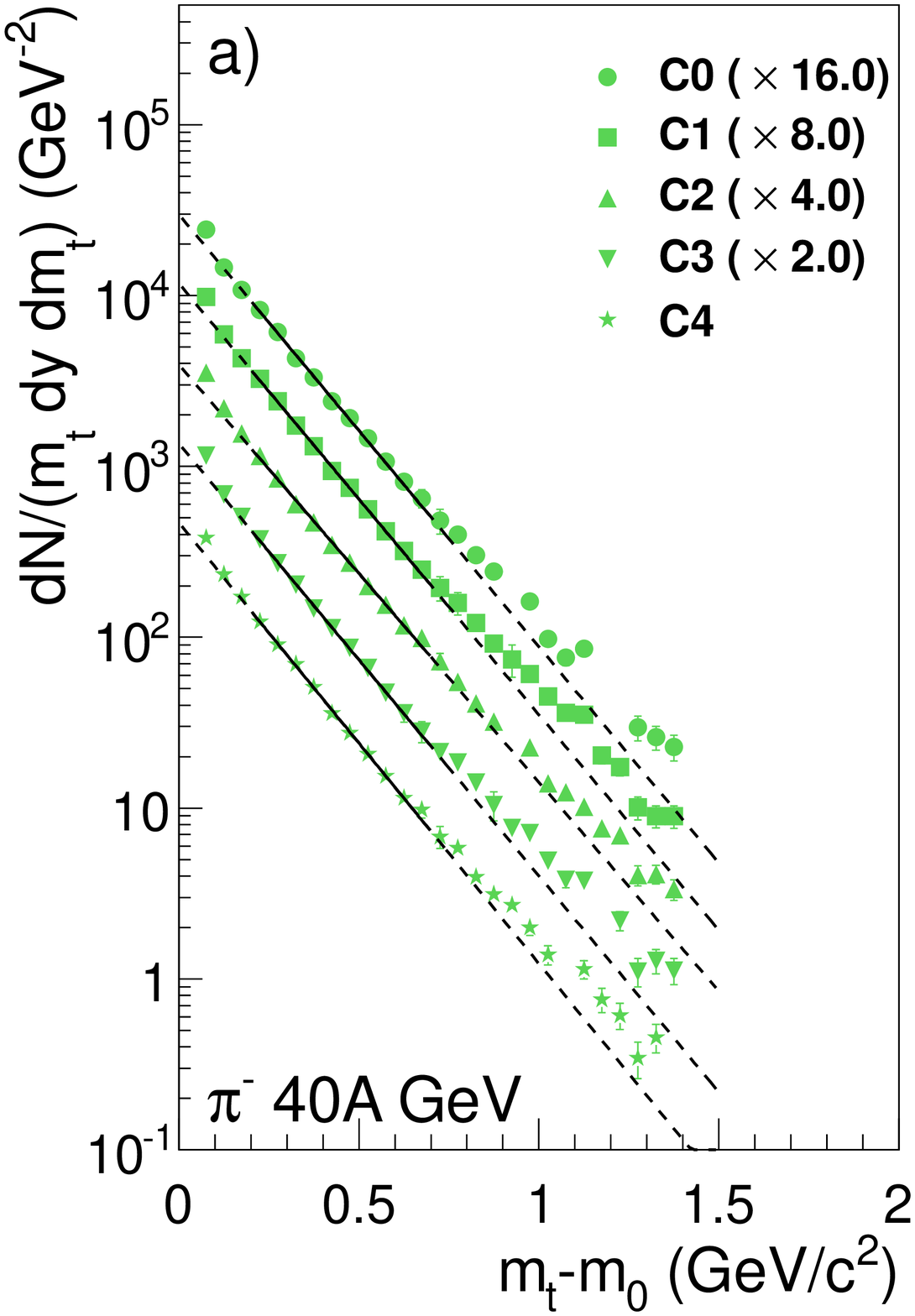}
 \hfill
 \includegraphics[width=0.44\textwidth]{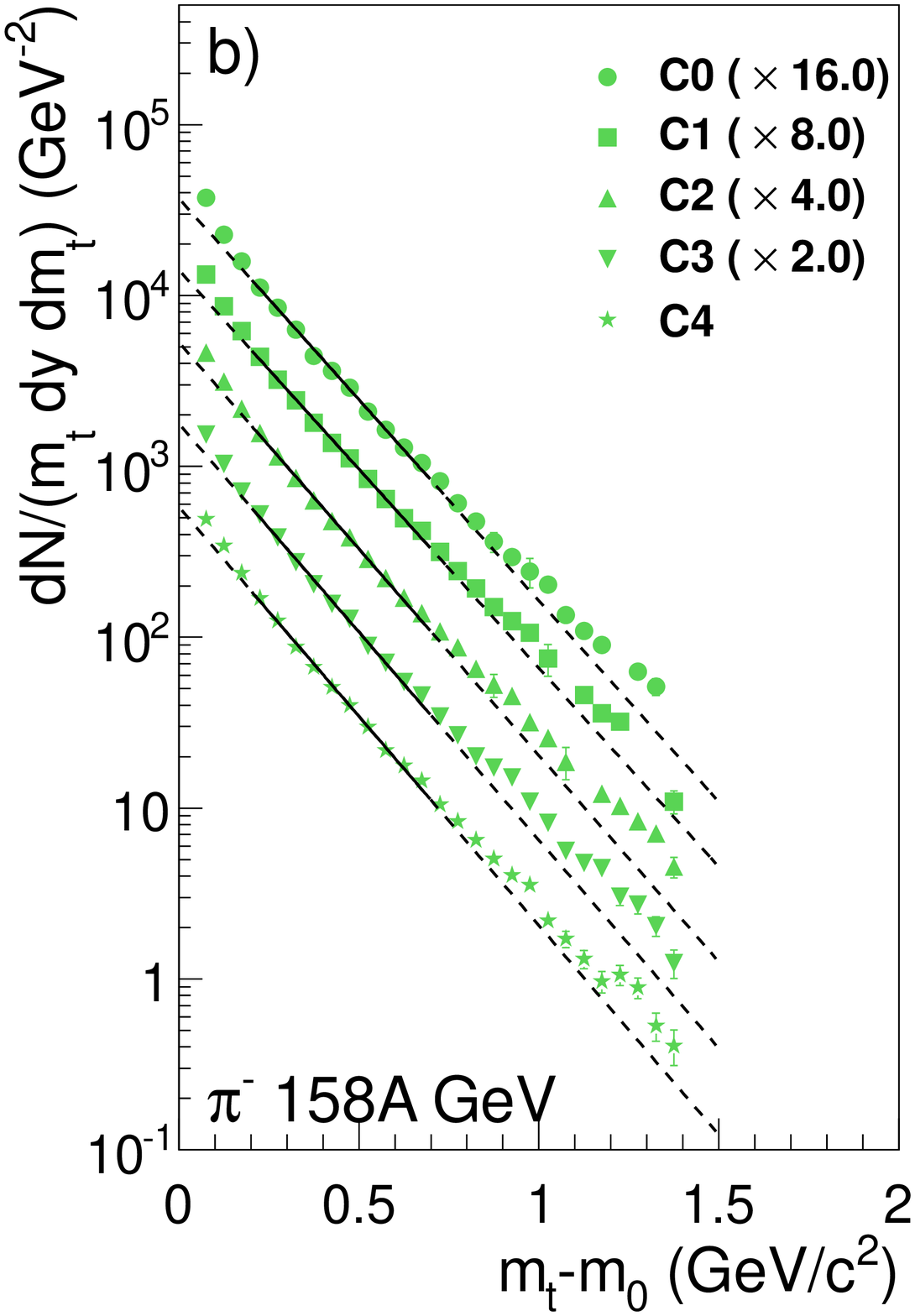}
\end{minipage}
\end{center}
 \vspace{-0.2cm}
\caption{\label{pt-spectra-pi}(Color online) Transverse mass spectra at
mid-rapidity ($0 < y < 0.2$) for $\pi^{-}$ in Pb+Pb collisions at 40\agev~and
158\agev~beam energy. Lines show exponential fits for
0.2~GeV/$c^{2}$ $< \mt-m_{0} <$ 0.7~GeV/$c^{2}$. Different
centralities are scaled for clarity. The error bars indicate the statistical
uncertainty.}
\end{figure}

\begin{figure}[h]
 \includegraphics[width=0.28\textwidth]{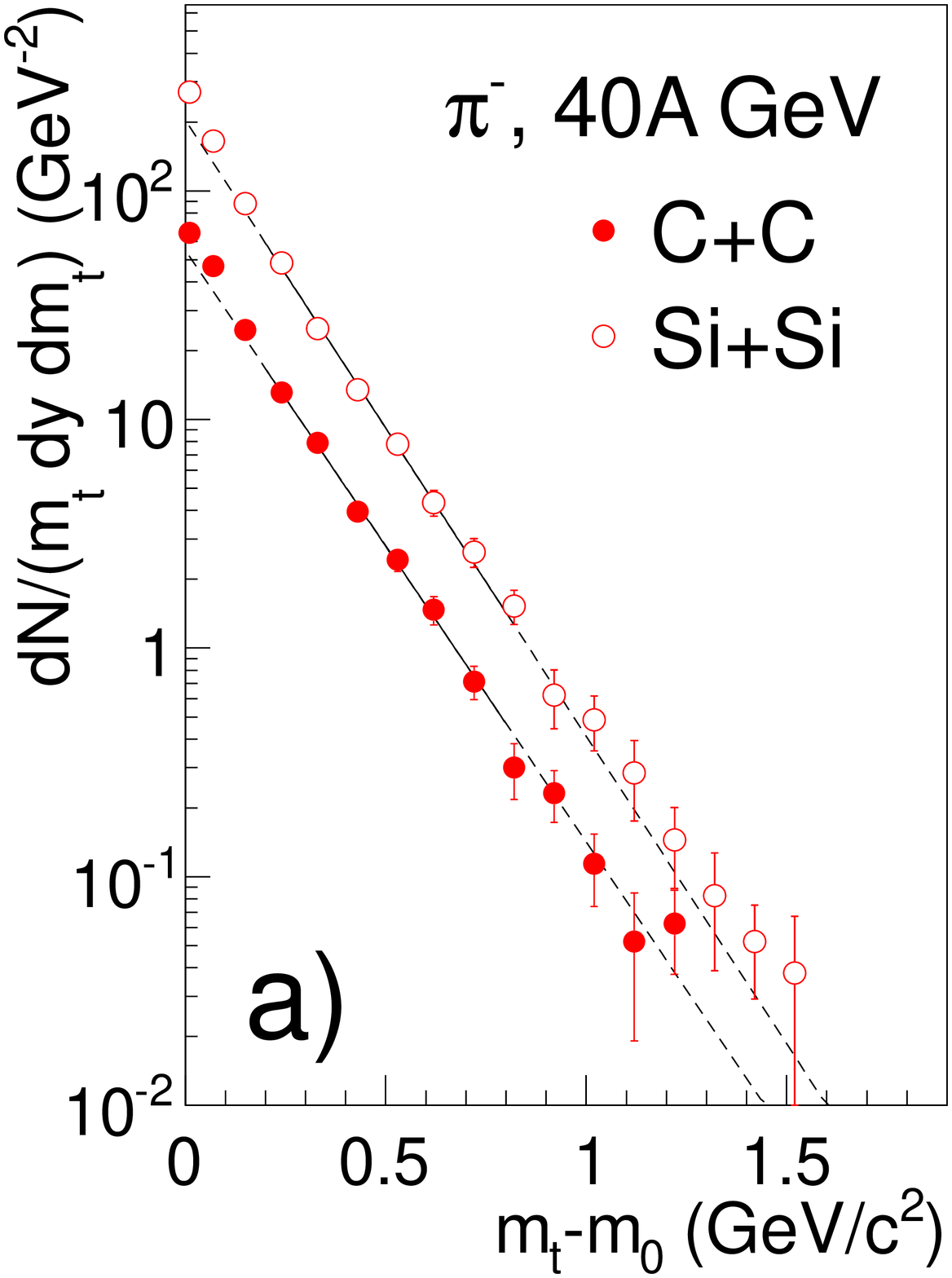}
 \includegraphics[width=0.28\textwidth]{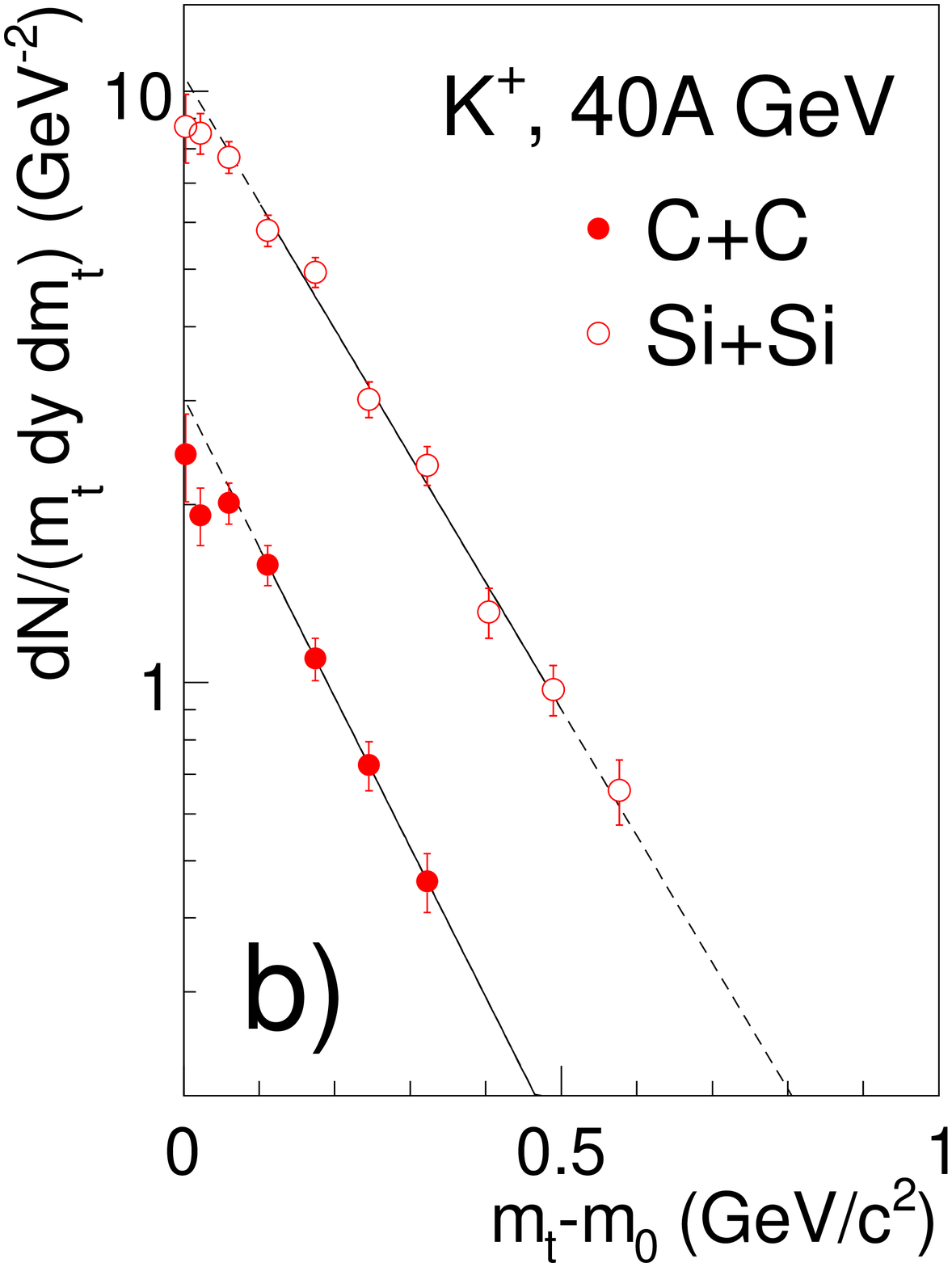}
 \includegraphics[width=0.28\textwidth]{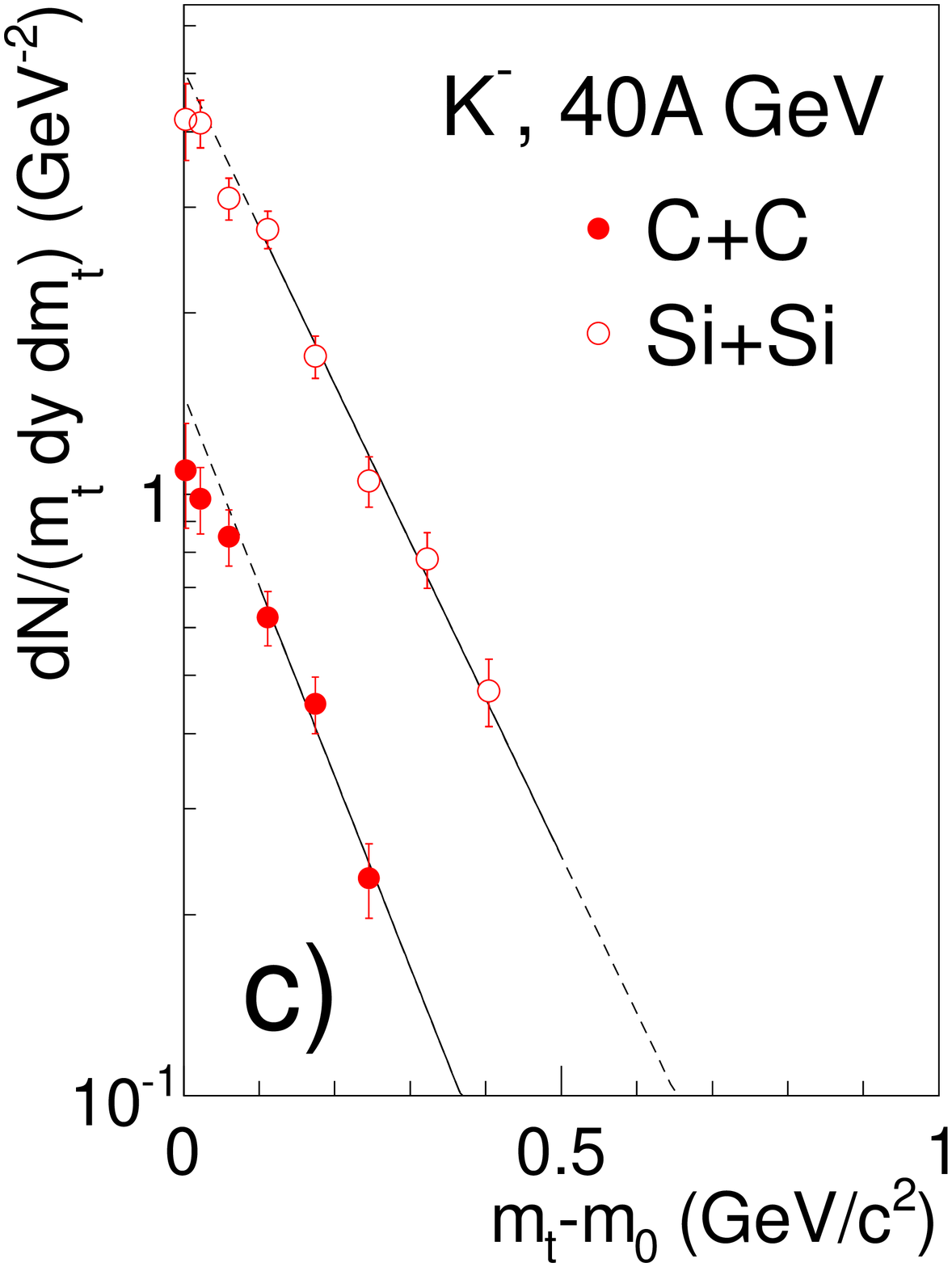}
 \vspace{-0.2cm}
\caption{\label{pt-spectra-cc-sisi}(Color online) Transverse mass spectra 
 of $\pi^{-}$ ($0 < y < 0.2$) and kaons ($0.8 < y < 1.0$)
at 40\agev~in C+C and Si+Si collisions. Lines show exponential
fits for 0.2~GeV/$c^{2} < \mt-m_{0} <$ 0.8~GeV/$c^{2}$  ($\pi^{-}$) 
and 0.1~GeV/$c^{2} < \mt-m_{0} < 0.5$~GeV/$c^{2}$ (kaons). The 
inverse slope
parameters for \kplus (\kmin) are  171 $\pm$18 MeV for C+C and 203 $\pm$10 MeV 
for Si+Si collisions (137 $\pm$24 MeV for C+C and 166 $\pm$13 MeV for Si+Si
collisions). The inverse slopes of pions are given in Table~\ref{CC-SiSi-40}.
The error bars indicate the statistical uncertainty.}
\end{figure}

\begin{figure}[]
 \center\includegraphics[width=0.44\textwidth]{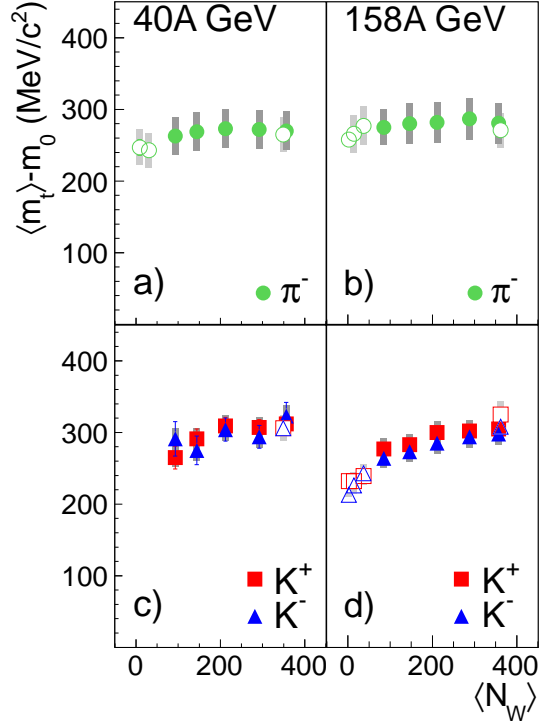}
 \vspace{-0.2cm}
\caption{\label{pt-spectra-mean}(Color online) Mean transverse mass as extracted
from the transverse mass spectra at mid-rapidity for $\pi^{-}$, K$^{+}$ and
K$^{-}$ versus number of participating nucleons. Open symbols show
results from central p+p, C+C, Si+Si and Pb+Pb collisions. Results from C+C and
Si+Si collisions at 40\agev~beam energy are new. The p+p data are extracted from
 \protect\cite{Fischer_pi,Fischer_K}. All other data points marked by open
symbols are from \protect\cite{na49_size,na49_energy,na49_20-30}. The thin 
vertical bars indicate the statistical errors which are mostly smaller than the
symbols. The thick shaded bars indicate the systematic errors.}
\end{figure}

Transverse mass spectra at mid-rapidity are shown in
Figs.~\ref{pt-spectra-K} and \ref{pt-spectra-pi} for kaons
(combined \dedx\ and TOF analysis) and pions for different
centrality classes in Pb+Pb collisions at 40\agev~and 158\agev~beam energy.
Fig.~\ref{pt-spectra-cc-sisi} presents results for
kaons and pions near mid-rapidity in C+C and Si+Si collisions at 40\agev. The
\mt~spectra (\mt~$= \sqrt{(m^2 + \pt^2)}$) were fitted by an exponential 
function with inverse slope $T$
\begin{equation}
\frac{\mbox{d}N}{\mt\mbox{d}\mt~dy}=C\cdot \exp{\left(-\frac{\mt}{T}\right)}
\end{equation}
in $\mt-m_{0}$ ranges as given in the figure
captions. In Pb+Pb collisions the transverse mass spectra of kaons are well 
described by this functional form, while pion spectra deviate at high and low 
transverse masses. The \mt~spectra of kaons in the light systems exhibit a 
two to three times statistical error downward deviation at low \mt.
For a model independent study of the transverse
mass spectra, the average transverse mass $\langle \mt~\rangle -
m_{0}$ was therefore calculated. To account for the small
unmeasured high \mt~part of the kaon spectra, which is on a percent level only, 
the spectra were extrapolated by the exponential functions shown in
Figs.~\ref{pt-spectra-K} and \ref{pt-spectra-cc-sisi}. For the pion spectra 
in Pb+Pb collisions an
exponential function with two slope parameters was used (not shown), to account
for the concave shape of the distributions (in logarithmic representation).
An estimate of systematic errors was derived from using different fit ranges
for the extrapolation or a single exponential for pions. The
resulting average transverse masses and inverse slope parameters
are summarized in Tables \ref{PbPb-158}, \ref{PbPb-40}, and
\ref{CC-SiSi-40}. Figure~\ref{pt-spectra-mean} shows the dependence
of $\langle \mt \rangle - m_{0}$ on the mean number of wounded
nucleons $\nwound$ for the investigated collision
systems at 40\agev~and 158\agev~beam energy. 
Results from central Pb+Pb \cite{na49_energy,na49_20-30}, p+p at 158 GeV/$c$
\cite{Fischer_K}, C+C and Si+Si collisions \cite{na49_size} at 
158\agev\ are also shown.

Owing to the mass difference and well known radial flow effect,
$\langle \mt\rangle - m_{0}$ values are larger for kaons than
for pions. For central collisions
\cite{na49_energy,na49_20-30} mean transverse masses do not change 
significantly from
40\agev~to 158\agev. For pions, if any, only a weak centrality
dependence is observed at 40\agev~and none at 158\agev~beam
energy, while kaons show an increase of mean transverse masses
towards central Pb+Pb collisions. This increase is particularly
pronounced when comparing to the much smaller C+C and Si+Si
collision systems with $\nwound < 60$.

\subsection{Rapidity spectra}
\begin{figure*}[]
 \center\includegraphics[width=0.9\textwidth]{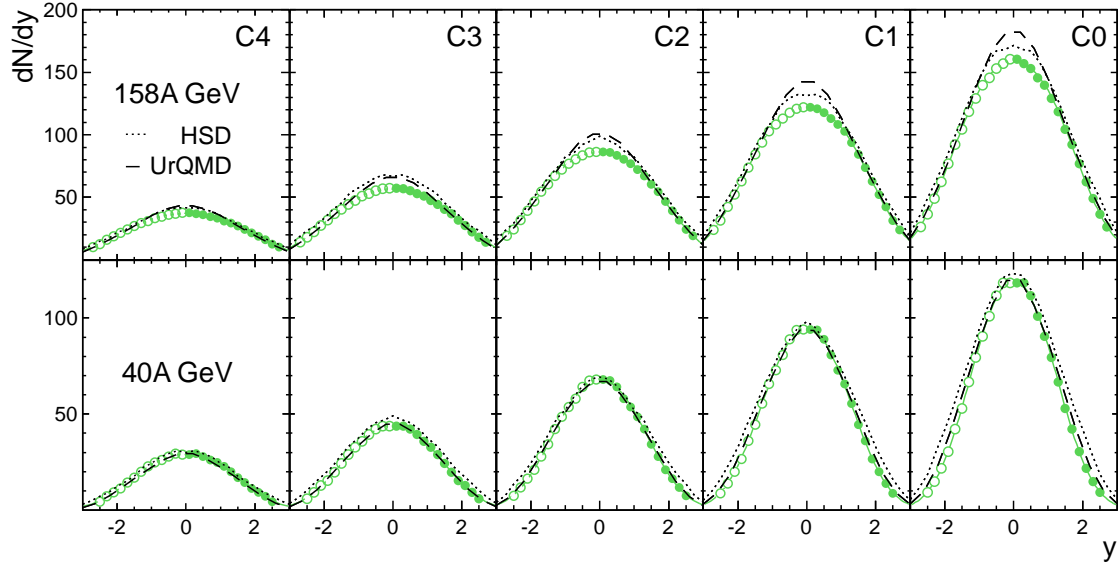}
 \vspace{-0.2cm}
\caption{\label{y-spectra-pi}(Color online) Rapidity spectra for $\pi^{-}$ 
in Pb+Pb collisions at 
40\agev~and 158\agev~beam energy for different centrality bins.
Open symbols are reflected at mid-rapidity.
Solid lines represent the fits of Eq.~\ref{EQ_dndy}.
The dotted and dashed lines represent HSD \cite{HSD} and UrQMD2.3 \cite{UrQMD} 
simulations, respectively. The error bars (mostly not visible) indicate the statistical
uncertainty.}
\end{figure*}
\begin{figure*}[]
 \center\includegraphics[width=0.9\textwidth]{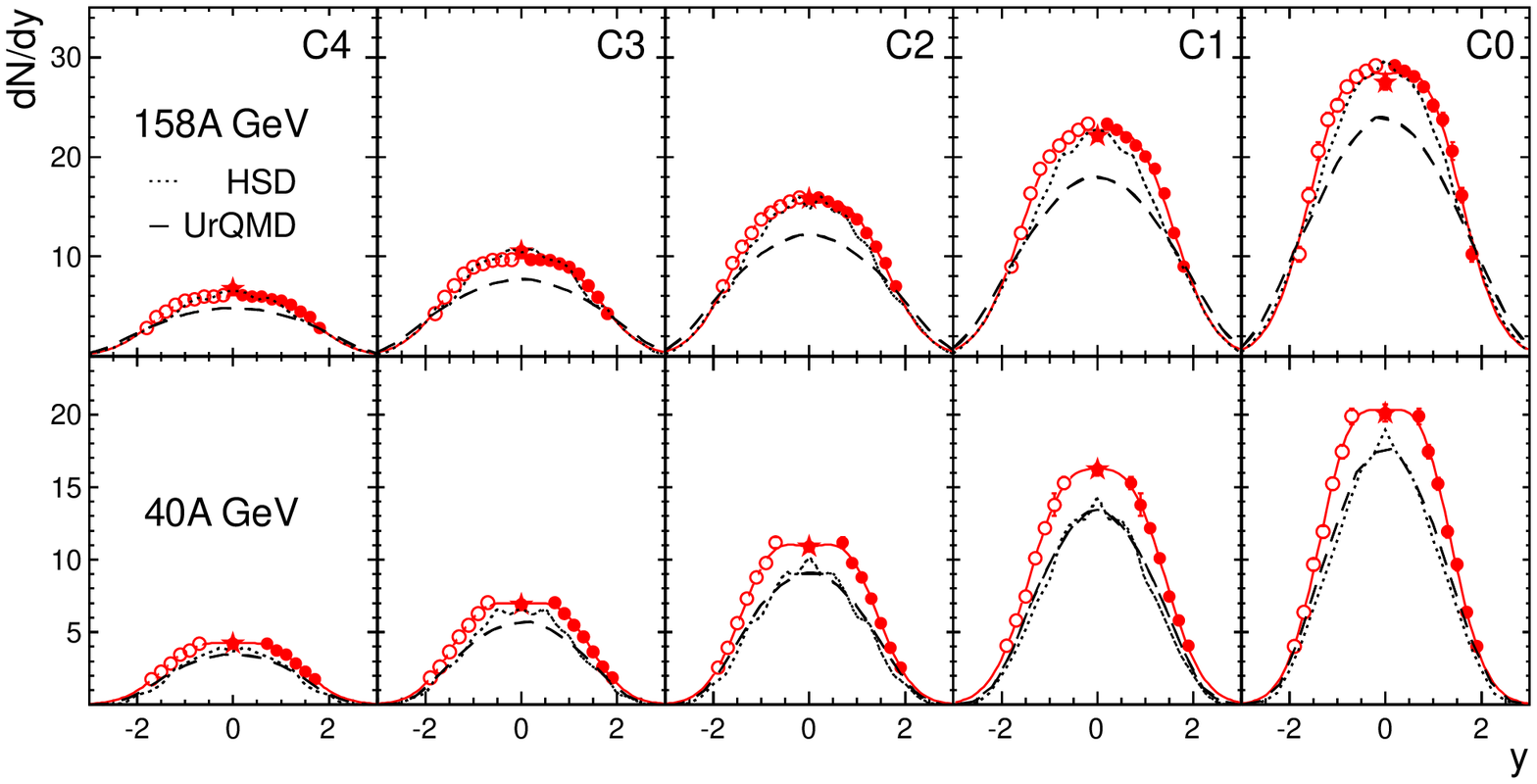}
 \vspace{-0.2cm}
\caption{\label{y-spectra-Kp}(Color online) Rapidity spectra of K$^{+}$ in Pb+Pb
collisions at 40\agev~and 158\agev~beam energy for different centrality bins.
Circles are values from the \dedx\ analysis, the stars give the
result from the combined \dedx\ and TOF analyses. Open symbols
are reflected at mid-rapidity. Solid lines represent the fits of Eq.~\ref{EQ_dndy}.
The dotted and dashed lines represent
HSD \cite{HSD} and UrQMD2.3 \cite{UrQMD} simulations, respectively.
The error bars (mostly not visible) indicate the statistical uncertainty.}
\end{figure*}
\begin{figure*}[hb]
 \center\includegraphics[width=0.9\textwidth]{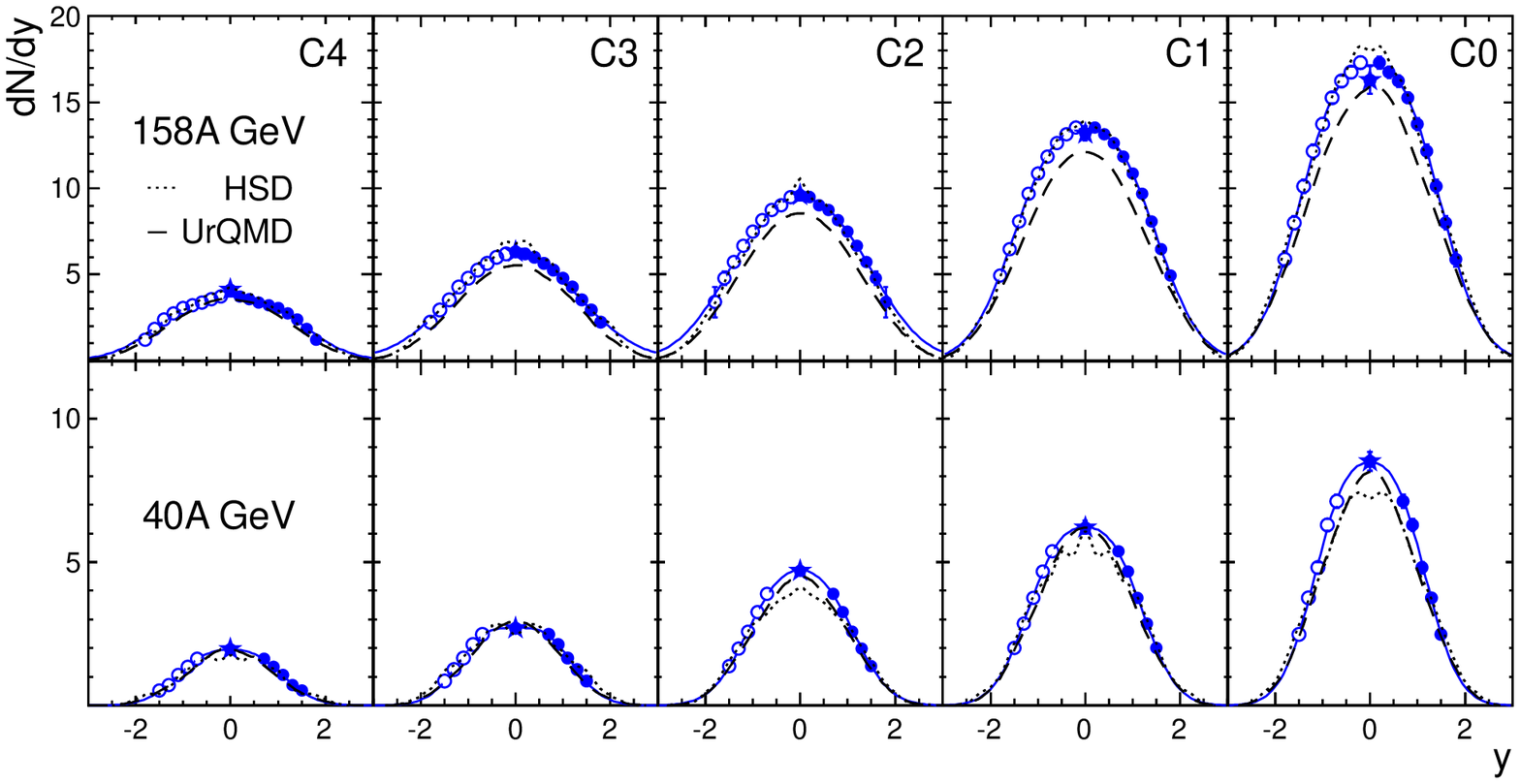}
 \vspace{-0.2cm}
\caption{\label{y-spectra-Km}(Color online) Rapidity spectra of K$^-$ in Pb+Pb collisions at
40\agev and 158\agev~beam energy for different centrality selections.
Circles are values from the \dedx\ analysis, the star gives the
result from the combined \dedx\ and TOF analysis. Open symbols
are reflected at mid-rapidity. Solid lines represent the fits of Eq.~\ref{EQ_dndy}. 
The dotted and dashed lines
represent HSD \cite{HSD} and UrQMD2.3 \cite{UrQMD} simulations, respectively.
The error bars (mostly not visible) indicate the statistical uncertainty..}
\end{figure*}
\begin{figure}[]
 \center\includegraphics[width=0.45\textwidth]
{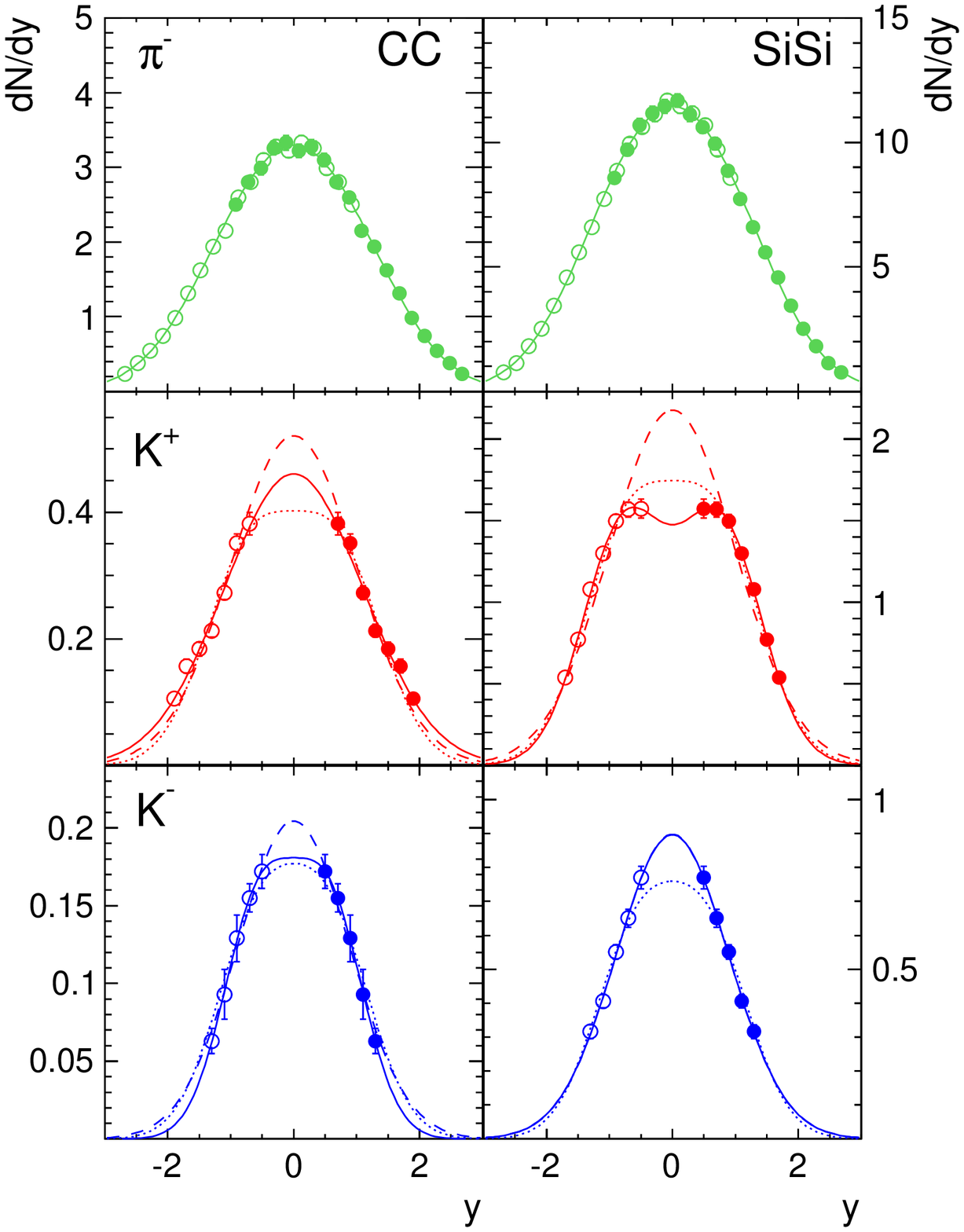}
 \vspace{-0.2cm}
\caption{\label{y-spectra-40}(Color online) Rapidity spectra for $\pi^{-}$ and
K$^{\pm}$ for semi-central C+C and Si+Si collisions at 40\agev~beam energy.
Solid lines show free fits of double Gaussians. Open symbols
are reflected at mid-rapidity. For
K$^{\pm}$ also double Gaussians with width $\sigma$ and shift
$y_{0}$ as for central Pb+Pb collisions at 40\agev~beam energy
were fitted (short dashed line) as well as a single Gaussian with
the same $RMS_{y}$ (dashed line). The error bars indicate the statistical
uncertainty.}
\end{figure}
Transverse momentum spectra were measured in bins of rapidity
in order to extract the rapidity dependence of \pt~-integrated
yields. The measured \pt~-spectra were extrapolated into
unmeasured regions by a single exponential for both kaons and 
pions in C+C and Si+Si interactions. A double exponential had to be used
for pions in Pb+Pb collisions because of the
concave shape of their \pt~-spectra. Rapidity distributions
of \pt~-integrated yields for the particles, collision
systems and energies under study are shown in
Figs.~\ref{y-spectra-pi}, \ref{y-spectra-Kp}, \ref{y-spectra-Km},
and \ref{y-spectra-40}. They are well described by two Gaussians
of equal widths $\sigma$ which are displaced symmetrically around
mid-rapidity by a constant $y_{0}$:
\begin{equation}
\frac{\mbox{d}N}{\mbox{d}y}=
C\left[\exp{\left(-\frac{(y-y_{0})^{2}}{2\sigma_{y}^{2}}\right)}+
                     \exp{\left(-\frac{(y+y_{0})^{2}}{2\sigma^{2}}\right)}
                     \right].
\label{EQ_dndy}
\end{equation}
Using this functional form the measured rapidity spectra can be
extrapolated into the unmeasured regions, and particle yields in
full phase space can be extracted. 
For C+C and Si+Si collisions at
40\agev~beam energy no combined \dedx\ and TOF measurement is
available at mid-rapidity. The resulting uncertainties in the shapes and integrals of the
rapidity distributions of kaons were evaluated by applying three different fit 
functions (Eq.~\ref{EQ_dndy}): with two free parameters ($\sigma_{y}$ and $y_0$), 
with these parameters fixed to values obtained  from fits to the corresponding 
distributions from
central Pb+Pb collsions at 40\agev, and with a single Gaussian  with  
its root mean square (RMS) again taken
from the 40\agev~Pb+Pb data. These fits are indicated in
Fig.~\ref{y-spectra-40} by the solid, short-dashed and dashed
lines and were used to extrapolate into the unmeasured
rapidity regions. Differences in the total yields between extremes are not exceeding 12\%
and are included in the systematic errors. The quoted total and mid-rapidity yields 
are the means of the values obtained from the three fits.
\begin{figure}[]
 \center\includegraphics[width=0.66\textwidth]{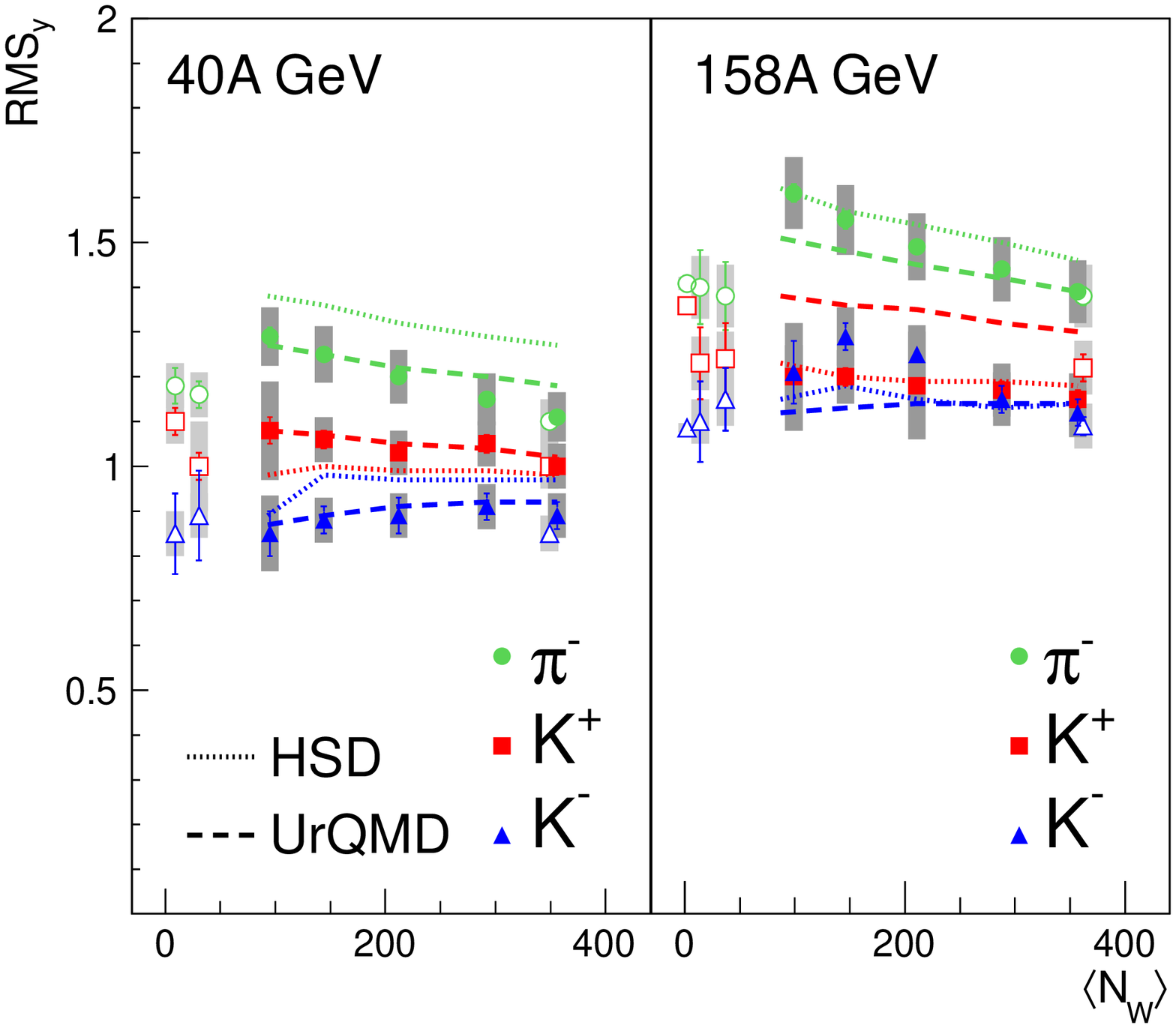}
 \vspace{-0.2cm}
\caption{\label{y-spectra-rms}(Color online) Width of the rapidity distribution
described by the RMS values for $\pi^{-}$, K$^{+}$ and K$^{-}$
versus mean number of participating nucleons. Open symbols show results
from p+p, C+C, Si+Si and central Pb+Pb collisions. Except for the new
results for C+C and Si+Si collisions at 40\agev\ beam energy
the data are taken from \protect\cite{Fischer_pi,na49_size,na49_energy,
na49_20-30}. Filled symbols represent the centrality selected Pb+Pb data.
Lines show results from HSD \cite{HSD} and UrQMD2.3 \cite{UrQMD} 
simulations as indicated in the figure. The thin 
vertical bars indicate the statistical errors, which are mostly not visible. 
The thick shaded bars indicate the systematic errors.}
\end{figure}
The widths of the rapidity distributions are quantified by
$RMS_{y} = \sqrt{\sigma_{y}^2+y_0^2}$. These values are tabulated together 
with mid-rapidity
and full phase space yields in Tables \ref{PbPb-158},
\ref{PbPb-40} and \ref{CC-SiSi-40}. Figure~\ref{y-spectra-rms}
presents the system-size dependence of the widths of the rapidity
distributions. While kaons show no significant change of the width
as a function of the  number of wounded nucleons, rapidity
distributions of pions are significantly wider in peripheral 
than in central Pb+Pb collisions. In fact, it turns out that
the  width of the pion rapidity distribution in central Pb+Pb
collisions is the same as in C+C and Si+Si reactions
as well as in p+p interactions at 158 \gevc.
Only non-central Pb+Pb collisions deviate from this common behavior exhibiting
a wider distribution. The widening of the pion rapidity distribution for
peripheral collisions is probably due to pion production in interactions of 
participants with spectator matter which had been already invoked to explain 
the proton rapidity spectra in the same data sets \cite{Milica}.

\subsection{Particle yields}

\begin{figure}[ht]
\includegraphics[width=0.7\linewidth]{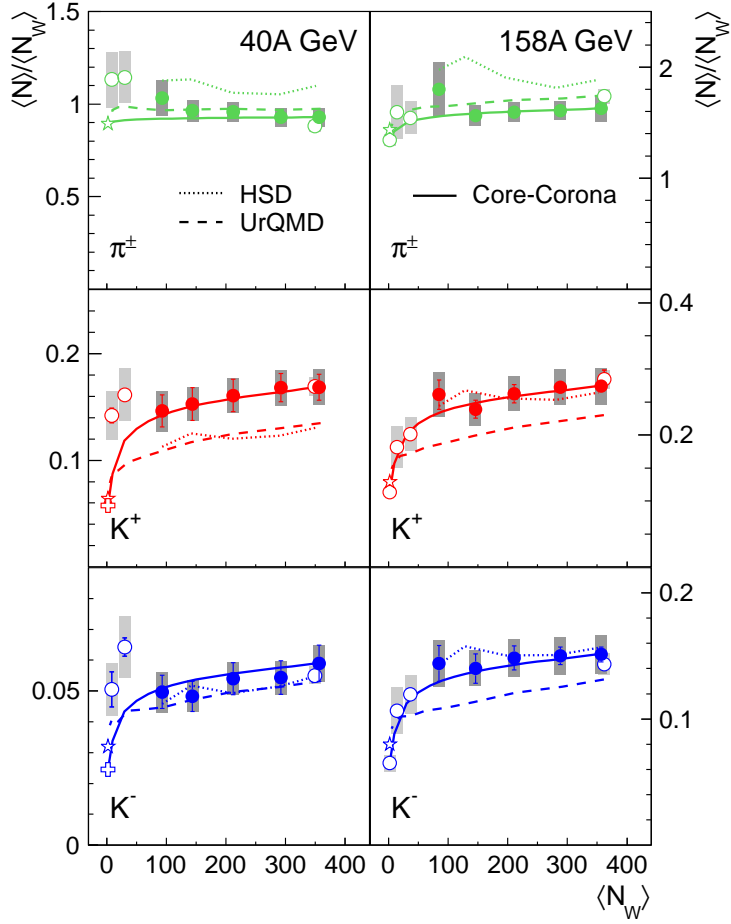}
 \vspace{-0.2cm}
\caption{\label{yield-nw}(Color online) Particle yields divided by the 
mean number of wounded nucleons as a function of centrality. Closed symbols show
results from centrality selected Pb+Pb collisions, open circles
 p+p data from NA49 \protect\cite{Fischer_pi,Fischer_K} as well as C+C, Si+Si 
(at 158\agev~\protect\cite{na49_size}) and central Pb+Pb collisions from
\protect\cite{na49_energy,na49_20-30}). The stars denote the values obtained 
for p+p interactions from the parametrization "B" in reference \protect\cite{rossi}. 
At 40 \gevc\ they are compared to values extracted from Fig. 132 reported
in \protect\cite{Fischer_K}.
Lines show results from HSD \cite{HSD} and UrQMD2.3 \cite{UrQMD}  calculations 
as well as from the core-corona model (the latter for the Pb+Pb collision 
system only, \protect\cite{blume}).
The thin  vertical bars indicate the statistical errors, which are mostly not visible. 
The thick shaded bars indicate the systematic errors.}
\end{figure}

\begin{figure}[]
\includegraphics[width=0.7\linewidth]{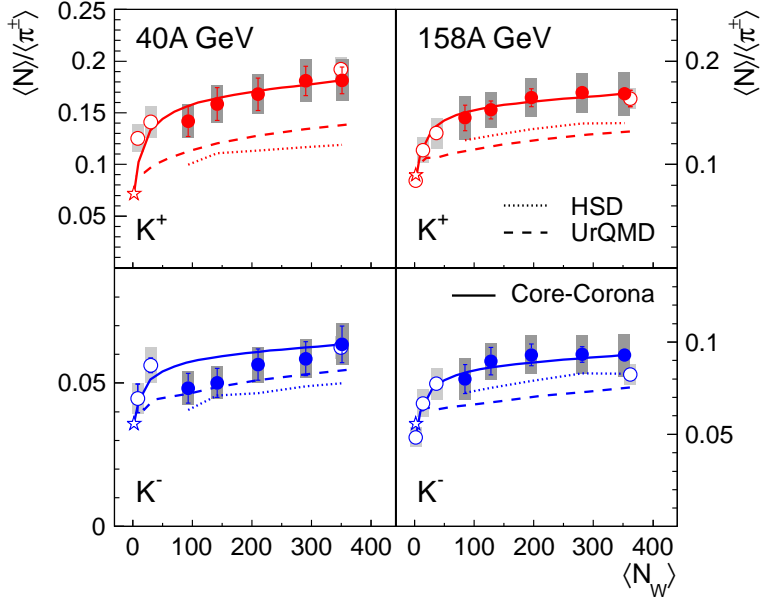}
 \vspace{-0.2cm}
\caption{\label{yield-pi}(Color online) Kaon yields divided by mean number
of pions $\langle \pi^{\pm} \rangle = 0.5\cdot\left(\langle
\pi^{-} \rangle + \langle \pi^{+} \rangle \right)$ as a function of centrality.
Closed symbols show
results from centrality selected Pb+Pb collisions, open symbols
 p+p data from NA49 \protect\cite{Fischer_pi,Fischer_K} as well as C+C, Si+Si 
(at 158\agev~\protect\cite{na49_size}) and central Pb+Pb collisions from
\protect\cite{na49_energy,na49_20-30}). The stars denote the values obtained 
from the parametrization "B" in reference \protect\cite{rossi}. Lines show
results from HSD \cite{HSD} and UrQMD2.3 \cite{UrQMD} simulations and the 
core-corona model (the latter for the Pb+Pb collision system only,\protect\cite{blume}).
The thin  vertical bars indicate the statistical errors, which are mostly not visible. 
The thick shaded bars indicate the systematic errors.}
\end{figure}

Fig.~\ref{yield-nw} shows the system-size dependence of particle
yields normalized to the mean number of wounded nucleons. Earlier data
on central C+C, Si+Si and Pb+Pb collisions are plotted as well
\cite{na49_size,na49_energy,na49_20-30}. Data for p+p collisions are taken
from NA49 where available \cite{Fischer_pi,Fischer_K} and from 
parametrizations by Rossi et al.~\cite{rossi}. At 40 \gevc\ we include
results from a new study of the energy dependence of kaon production reported
in \cite{Fischer_K}. Also shown are calculations 
by the HSD \cite{HSD} and UrQMD2.3 \cite{UrQMD} transport models, as well as 
by the core-corona model (the latter for the Pb+Pb collision system only, see also
Fig.~\ref{corcor}) \cite{blume}. Normalized pion yields
are rather independent of centrality or system size while the
normalized yields of kaons show a steep increase for small system
sizes followed by a slow rise or even saturation for higher centralities. 
The mean number of wounded nucleons
($\nwound$) introduces a model dependence in these
ratios, in particular for small systems, as for the semi-central C+C and
Si+Si collisions the calculations have large uncertainties. Yields
are therefore alternatively normalized to the number of pions. 
In Fig.~\ref{yield-pi}
these ratios are presented. Normalized kaon yields at both
energies exhibit the same dependence as seen in Fig.~\ref{yield-nw}.
A closer look reveals that at 158\agev~the  K/$\pi$ ratios in the small 
collision systems C+C and Si+Si seem to follow the trend set by the
centrality dependence of the Pb+Pb data, whereas at 40\agev~the 
normalized kaon 
abundance in peripheral Pb+Pb collisions is lower than expected from
the extrapolation of the K/$\pi$ ratio in C+C and Si+Si collisions.


\section{Discussion}

\subsection{Comparison to transport models (HSD and UrQMD)}
The centrality of the events generated by the models was derived from the 
corresponding fraction of the total (geometrical) cross section  with the
impact parameter as order parameter. The corresponding \nwound~values were  
those derived from the experimental data for HSD. For the UrQMD events the 
number of nucleons in the nuclear overlap volume \nwound~ was calculated 
event-by-event based on the chosen impact parameter.
Pion rapidity distributions (Fig.~\ref{y-spectra-pi}) are well reproduced by 
both models at 40\agev,
whereas their predictions give slightly higher yields than the data at 158\agev. 
For K$^+$ mesons (Fig.~\ref{y-spectra-Kp}) the
distributions from both models are 20\% low at 40\agev. For kaons 
at 158\agev~HSD calculations agree well with experiment, whereas UrQMD2.3 
underestimates our data.
The overall agreement betweeen model calculations and data for the K$^-$ mesons
(Fig.~\ref{y-spectra-Km}) is generally good for both models, even though 
there is a small underestimation of the data by UrQMD2.3 at 158\agev~and 
by HSD at 40\agev. No model data were provided for comparison to our C+C 
and Si+Si data. As to the shapes of the rapidity distributions 
both models describe well the increase of RMS$_y$ with increasing impact
parameter at both energies (Fig.~\ref{y-spectra-rms}). This increase is most
(least) pronounced for the pions (K$^-$). We interpret this rise as due to
interactions of fireball particles with spectator remnants. 

Normalized multiplicities are studied in Figs.~\ref{yield-nw} and
\ref{yield-pi}. $\langle N \rangle$/$\nwound$ for pions is
reproduced by UrQMD2.3 at both energies. HSD is 10\% high, which is due to larger
widths rather than higher amplitudes of the corresponding \dndy~distributions 
(cf. Fig.~\ref{y-spectra-pi}). We now turn to the normalized yields of the 
K$^+$ meson. At 40\agev~the results from both models are low by 20\%. At
158\agev~HSD results agree with our data, wheras UrQMD2.3 results underpredict
them. The K$^-$ yields are well desribed by both models
at 40\agev. At 158\agev~K$^-$ mesons yields are reproduced by HSD, whereas they
are underpredicted by UrQMD2.3 as was the case for K$^+$ yields. 
The calculations of both
models for the kaon to pion ratios yield values which are lower than the 
experimental data at both energies (see Fig.~\ref{yield-pi}), although
for different reasons. While UrQMD2.3 mainly underpredicts the kaon 
yields, HSD overpredicts the pions yields (see 
Figs.~\ref{yield-nw}). 

\subsection{Transverse mass distributions and thermal freeze-out}
Both the thermal freeze-out temperature and  radial flow affect the observed 
\mt~distributions of produced particles. The
centrality dependence of $\langle \mt \rangle-m_0$ (near mid-rapidity), which 
is shown in Fig.~\ref{pt-spectra-mean}, exhibits a clear trend of an increase 
of $\langle \mt \rangle-m_0$ with centrality (for Pb+Pb) and system size 
(p+p and light
versus heavy systems), which is most pronounced for kaons of both charges at
158\agev.  This finding is consistent with the picture that nuclear stopping and
thus energy deposition increases with increasing centrality and from
small to large collision systems, giving rise to higher temperatures or
collective radial flow or both.

\subsection{System size dependence of the K/$\pi$ ratio}

\begin{figure}[h]
\includegraphics[width=0.7\linewidth]{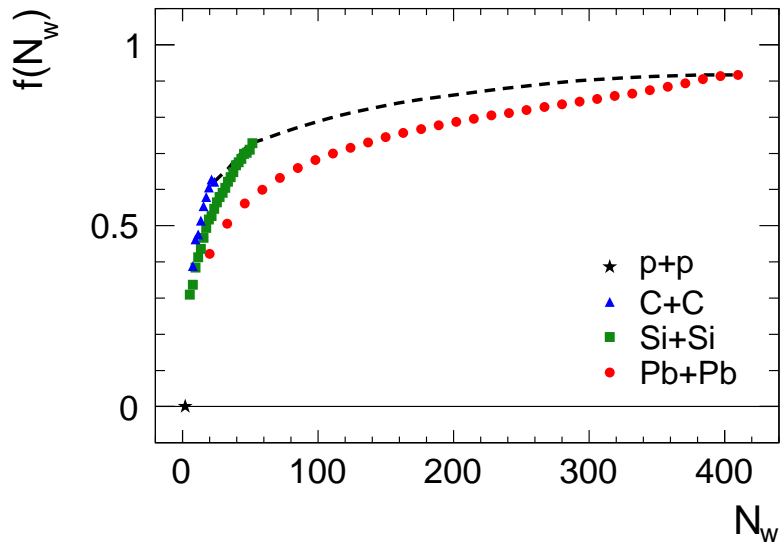}
 \vspace{-0.2cm}
\caption{\label{corcor}(Color online)
Core fraction obtained in the core-corona model for different collision 
systems as function of the number of 
participating nucleons \protect\cite{blume}. The dashed line connects the 
endpoints of
the curves for the different collision systems. $f(\nwound)$ denotes the 
fraction of those nucleons, which undergo more than one collision.}
\end{figure}
The system size dependence of the K/$\pi$ ratios 
(see  Fig.~\ref{yield-pi}) hints at a behavior expected for the core-corona 
ansatz \cite{blume}. However at 40\agev~there are significant deviations of the 
core-corona calculations from the experimental data. In peripheral Pb+Pb
collisions the data tend to be below the model values,
most pronounced for K$^-$, and the ratios measured for the small
collision system (C+C and Si+Si) tend to be above the values
calculated for the Pb+Pb system. The latter finding is in line with
the model prediction for different system sizes as shown in Fig.~\ref{corcor}.
It shows the core fraction obtained in the core-corona model for different 
collisions systems as function of the number of participating 
nucleons \cite{blume}.
Central collisions of small projectiles reach their maximum already for smaller 
volumes (number of wounded nucleons) than heavy systems like Pb+Pb. 
At 158\agev\ (see right column of Fig.~\ref{yield-pi}) the data points from 
peripheral Pb+Pb collisions follow nicely the core-corona prediction for the 
Pb+Pb system. Also the C+C and Si+Si points are consistent with this prediction
(for Pb+Pb), although according to Fig.~\ref{corcor} they should be higher even
for non-central collisions of the small C+C and Si+Si systems. However, here
the core-corona approach is applied to small system sizes 
for which the simplification of separating the collision zone into one
central cluster and p+p collisions in the periphery might not be
valid anymore. Instead, percolation calculations suggest that for
such collision systems several clusters of smaller size will be
formed \cite{percolation}.

\begin{figure}[h]
\includegraphics[width=0.7\linewidth]{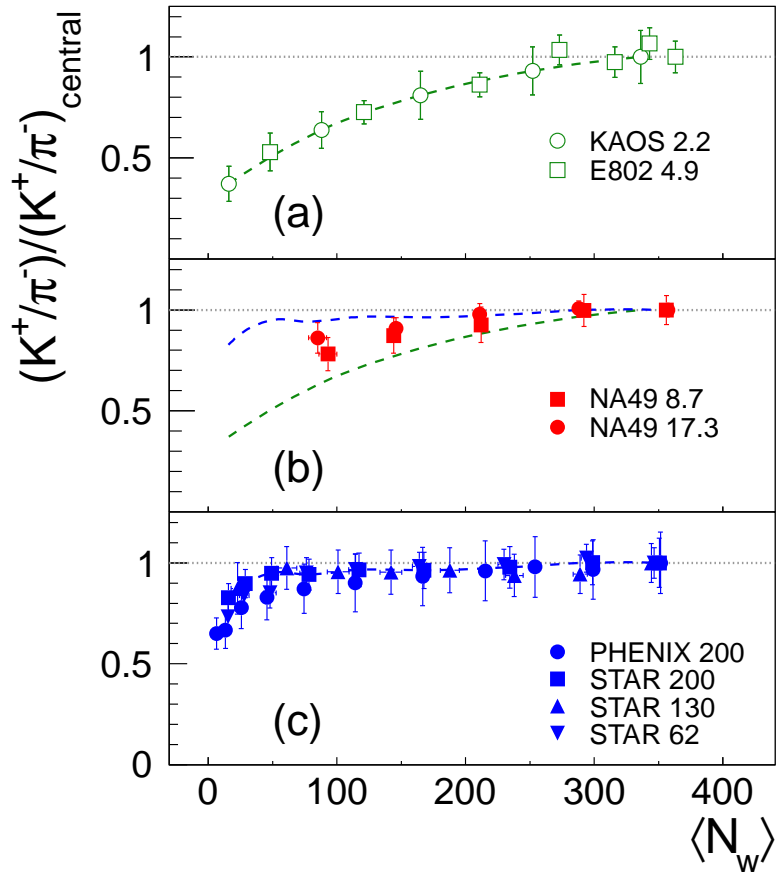}
 \vspace{-0.2cm}
\caption{\label{sqrts_K2pi}(Color online) Centrality dependence of the double
ratio of the K$^+$ to $\pi^-$ multiplicities  scaled to the  ratios from
the corresponding most central measurement 
minimum bias Pb+Pb (Au+Au) collisions at SIS \cite{sis} and AGS \cite{ags,Wang}
energies (a), at SPS energies (b) and at RHIC energies \cite{phenix,star} (c).
The dashed lines in panels (a) and (c) are meant to guide the eye and are 
also drawn as a reference in panel (b). Only statistical errors are shown.}
\end{figure}

In Fig.~\ref{sqrts_K2pi} we summarize the K$^+$/$\pi^-$ ratios for 
centrality selected Au+Au (Pb+Pb) collisions at different beam 
energies.  The ratios at the different centralities were scaled to the ratios
measured in the most central collisions. The observed saturation
of strangeness production at SPS energy and above can be understood
in the context of statistical models by the approach to a
grand-canonical description. Increasing the energy available for
particle production does no longer change the multiplicities of
strangeness carrying particles relative to pions \cite{na49_energy,na49_20-30},
which means that strangeness becomes fully saturated. 
Saturation already sets in for rather peripheral collisions. The reduction of
relative strangeness production towards even smaller system sizes has 
been discussed in the framework of the core-corona model in the previous 
paragraph. An implementation of this ansatz into the statistical model and 
a detailed comparison of model calculations to RHIC data can be found in 
\cite{becattini}. In order to describe the data satisfactorily it is essential 
to use a volume dependence which is not just proportional to \nwound.
The situation is different at low (SIS) energies. Here 
the K$^+$/$\pi^-$ ratio does not saturate and its dependence on the number of
wounded nucleons is rather linear. In contrast to the situation at high energies,
statistical models can describe
the observed centrality dependence at low energy within a canonical 
ensemble and with a volume proportional to the number of
wounded nucleons $V~\propto~\nwound$ \cite{Cle98}. Such a behavior would also be
expected in a hadronic rescattering scenario as realized in microscopic models
\cite{Wang}. The results obtained at 40\agev~are
intermediate between those at top SPS and AGS energies. We thus
observe a smooth transition between both scenarios which occurs 
in the SPS energy region.

\subsection{$E_{\mbox{s}}$ as function of system size}
\begin{figure}[]
\includegraphics[width=0.7\linewidth]{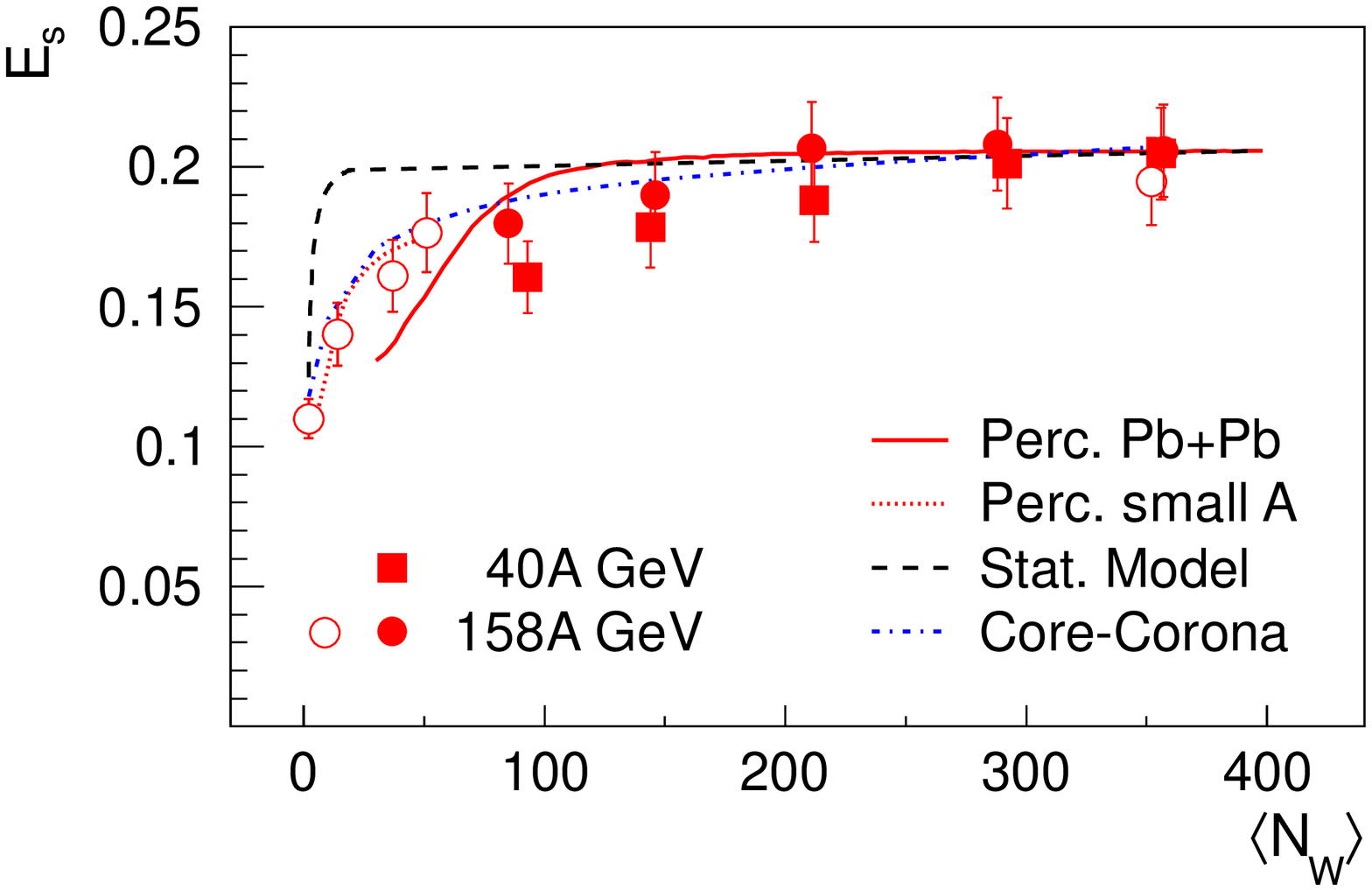}
 \vspace{-0.2cm}
\caption{\label{E_s}(Color online) Total relative strangeness approximated by 
$E_{\mbox{s}}$ for
centrality selected minimum bias Pb+Pb collisions at 40\agev~(squares) and 
158\agev~
(circles) as function of $\nwound$. The open symbols are from 
p+p \protect\cite{Fischer_pi,Fischer_K}, C+C,
Si+Si \protect\cite{na49_energy}, S+S \protect\protect\cite{Alber} and central
 Pb+Pb \protect\cite{na49_energy} at 158\agev. The filled symbols 
represent the data of centrality 
selected Pb+Pb collisions at 40\agev~and 158\agev~(this analysis 
and \cite{na49_multis}). The lines indicate calculations
within the statistical model assuming proportionality between
ensemble volume and  $\nwound$ (dashed), with volumes as 
derived from percolation calculations for small systems (dotted line) and 
Pb+Pb collisions (solid line) \protect\cite{percolation}, as well as 
calculations based on the core-corona ansatz at 158\agev~(dash-dotted line). 
Only statistical errors are shown.}
\end{figure}

The strangeness to entropy ratio or total relative strangeness
production described by the Wroblewski factor $\lambda_{s}={2
\langle s\overline{s} \rangle }/{\langle
u\overline{u}+d\overline{d}\rangle }$ can be approximated by the 
measurable quantity $E_{\mbox{s}}$= $(N_{\Lambda} + 2 [N_{K^+} 
+ N_{K^-}])/(1.5 [N_{\pi^+}+N_{\pi^-}])$ \cite{na49_energy}.
The data presented in this paper combined with earlier ones published in 
\cite{na49_multis,na49_size} allow to study the dependence of 
$E_{\mbox{s}}$ on system size for 40\agev~and 158\agev~beam energy with the 
158\agev~data covering the whole range of possible sizes. This 
dependence is shown in Fig.~\ref{E_s} and can now be compared directly to
the dependence obtained from statistical model calculations 
assuming for simplicity that $E_{\mbox{s}}~\propto~\eta$, with
$\eta$ being the canonical strangeness suppression factor, see for 
example Refs. \cite{rafelski-danos,redlich}. In statistical models the relative 
strangeness production depends on the volume of the ensemble. Assuming 
naively that the volume is proportional to the number of wounded nucleons
(\cite{rafelski-danos,redlich}) yields an increase of relative strangeness 
production (dashed line in Fig.~\ref{E_s}) which is significantly steeper 
than the increase observed in the data. On the
other hand, the smooth rise of  $E_{\mbox{s}}$ with system size and the
saturation around 60-100 wounded nucleons is reproduced within
experimental errors by a percolation model \cite{percolation}. 
In this model the volumes relevant for strangeness production by
statistical hadronization are calculated within a percolation
ansatz. This yields several smaller clusters in collisions of
light nuclei or in peripheral Pb+Pb collisions, and a large cluster accompanied
by small clusters (p+p interaction like) in the periphery for
central Pb+Pb collisions. The latter finding 
justifies the simplified ansatz used in the core-corona approach,
which is shown by the dash-dotted line. The
functional dependences change slightly between light
and heavy nuclei (dotted and solid lines in Fig.~\ref{E_s}), both
describing the data at 158\agev\ beam energy well. The Pb+Pb data 
show a weaker increase at 40\agev~than at 158\agev.

\section{Summary}
The NA49 collaboration measured the rapidity and transverse mass distributions 
of kaons and pions in Pb+Pb collisions at different centralities. These data 
are complemented 
by results from the small systems C+C and Si+Si. We find that the centrality 
dependence 
of kaon and pion production is reproduced by microscopic transport model 
calculations
(HSD and UrQMD2.3) within 20\% or better. The mean transverse mass evolves with 
centrality as 
expected for increased stopping and energy deposition. The centrality 
dependence of the
widths of the rapidity distributions in Pb+Pb collisions does not smoothly 
connect to 
results from the small systems
p+p, C+C, and Si+Si. We attribute this behavior to reinteractions of produced 
particles 
with spectator remnants, which is possible only in the heavy collision system. 
In Pb+Pb collisions the K/$\pi$ 
ratios show a smooth increase with centrality at both beam energies with 
saturation setting in
around 100-200 (60-100) wounded nucleons for 40\agev~(158\agev). 
In particular at 40\agev~beam energy the ratios measured in C+C and
Si+Si collisions tend to be higher than the ratios in peripheral
Pb+Pb interactions. The strong rise for small systems 
can be understood in the framework of statistical
models, if the relevant volume is not proportional to the number of
participants but is a superposition of relatively small
subvolumes for small system sizes and one large volume in central
collisions accompanied by p+p like interactions in the corona.

\newpage
\begin{table*}[h]
 \begin{ruledtabular}
  \begin{tabular}{ c c c c c c c}
  & Centr. & $\dndy$ & $\langle N \rangle$ & $RMS_{y}$ & $\langle \mt\rangle
-m_{0}$  & $T$ \\
  & class  &  &  &  &  MeV/c$^{2}$ & MeV \\
    \hline
 $\pi^{-}$ & 0 & 158.8 $\pm$0.7 $\pm$16 & 602 $\pm$4.7 $\pm$42
           & 1.39 $\pm$ 0.01 $\pm$0.07 & 281 $\pm$2 $\pm$28 & 184 $\pm$2 $\pm$9\\
           & 1 & 121.4 $\pm$0.6 $\pm$8.5 &  480 $\pm$4.5 $\pm$29
           & 1.44 $\pm$0.01 $\pm$0.07 & 287 $\pm$2 $\pm$28
           & 187 $\pm$1 $\pm$9\\
           & 2 & 86.2  $\pm$0.5 $\pm$4.3 & 349 $\pm$4.2 $\pm$17
           & 1.49 $\pm$0.01 $\pm$0.08 & 282 $\pm$2 $\pm$28
           & 180 $\pm$1 $\pm$9\\
           & 3 & 57.2  $\pm$0.4 $\pm$2.9 & 237 $\pm$3.8 $\pm$12
           & 1.55 $\pm$0.02 $\pm$0.08 & 280 $\pm$2 $\pm$28
           & 178 $\pm$2 $\pm$9\\
           & 4 & 37.2  $\pm$0.3 $\pm$1.9 & 159 $\pm$ 3.4 $\pm$7.9 
           & 1.61 $\pm$0.02 $\pm$0.08 & 275 $\pm$2$\pm$28
           & 177 $\pm$2 $\pm$9\\
    \hline
 K$^{+}$   & 0 & 28.35 $\pm$0.43$\pm$2.9& 97.8 $\pm$3 $\pm$9.8 & 1.15
        $\pm$0.02 $\pm$0.06 & 305 $\pm$6 $\pm$15 & 232 $\pm$3 $\pm$12\\
        & 1 & 22.72 $\pm$0.31 $\pm$2.3& 78.5 $\pm$2.8 $\pm$7.8 &
           1.17 $\pm$0.02 $\pm$0.06 & 302 $\pm$5 $\pm$15 & 229
           $\pm$3 $\pm$12\\
       & 2 & 15.86 $\pm$0.23 $\pm$1.6& 57.1 $\pm$2.9 $\pm$5.7 &
           1.18 $\pm$0.02 $\pm$0.06 & 300 $\pm$5 $\pm$15 & 227
           $\pm$2 $\pm$12\\
       & 3 & 10.01 $\pm$0.17 $\pm$1.0& 34.9 $\pm$2 $\pm$3.5 &1.21
           $\pm$0.02 $\pm$0.06 & 283 $\pm$6 $\pm$14 & 215
           $\pm$3 $\pm$11\\
       & 4 & 6.39 $\pm$0.13 $\pm$0.6& 23.2 $\pm$1.9 $\pm$2.3 &
           1.22 $\pm$0.02 $\pm$0.06 & 277 $\pm$7 $\pm$14 & 209
           $\pm$3 $\pm$11\\
    \hline
 K$^{-}$  & 0 & 16.8 $\pm$0.43 $\pm$1.7& 53.95 $\pm$2.15 $\pm$4.3& 1.12
       $\pm$0.03 $\pm$0.06& 298 $\pm$9 $\pm$15 & 225 $\pm$5 $\pm$11\\
       & 1 & 13.38  $\pm$0.24 $\pm$1.3& 43.18 $\pm$2.03 $\pm$3.5&
       1.15 $\pm$0.03 $\pm$0.06&  294 $\pm$7 $\pm$15 & 221 $\pm$4 $\pm$11\\
       & 2 & 9.54  $\pm$0.17 $\pm$0.9& 31.32 $\pm$2.03 $\pm$2.5& 1.25
       $\pm$0.1 $\pm$0.06&  285 $\pm$8 $\pm$14 & 217 $\pm$3 $\pm$11\\
       & 3 & 6.26  $\pm$0.12 $\pm$0.6& 20.45 $\pm$1.69 $\pm$1.6& 1.29
       $\pm$0.03 $\pm$0.07&  273 $\pm$8 $\pm$14 & 210 $\pm$4 $\pm$11\\
       & 4 & 3.92  $\pm$0.11 $\pm$0.4& 12.24 $\pm$1.18 $\pm$1.2& 1.21
       $\pm$0.07 $\pm$0.06&  264 $\pm$10 $\pm$13 & 202 $\pm$5 $\pm$10
\\
  \end{tabular}
 \end{ruledtabular}
\vspace{-0.2cm}
 \caption
 {\label{PbPb-158} Summary of data for pion and kaon production in centrality
selected Pb+Pb collisions at 158\agev~beam energy:
 rapidity density \dndy\ at mid-rapidity ($\pi^{-}$: $0<y<0.4$, K$^{\pm}$:
average of result from TOF analysis ($-0.2 < y < 0.2$) and \dedx\ analysis
 ($0.1<y<0.3$)), total multiplicity $\langle N \rangle$, RMS width of the
 rapidity distribution $RMS_{y}$, average transverse mass at mid-rapidity
$\langle \mt\rangle-m_{0}$, inverse slope parameter of \mt~spectra at
 mid-rapidity
 ($\pi^{-}$: $0<y<0.2$, fitrange 0.2 $<\mt-m_{0}<$ 0.7
 GeV/$c^{2}$;
  K$^{\pm}$: $-0.2<y<0.2$ (TOF analysis), fit range 0 $<\mt-m_{0}<$ 1.0
 GeV/$c^{2}$). The first error is statistical, the second
 systematic.}
\end{table*}

\newpage
\begin{table*}[t]
 \begin{ruledtabular}
  \begin{tabular}{ c c c c c c c}
  & Centr. & \dndy\ & $\langle N \rangle$ & $RMS_{y}$ & $\langle \mt\rangle
-m_{0}$  & $T$ \\
 & class  &  &  &  &  MeV/c$^{2}$ & MeV \\
    \hline
 $\pi^{-}$ & 0 & 118 $\pm$0.6 $\pm$12& 348 $\pm$2.4 $\pm$17.4 & 1.11 $\pm$0.01
           $\pm$0.06 & 270 $\pm$3 $\pm$27 & 172 $\pm$3 $\pm$9\\
           & 1 & 93.8 $\pm$0.5 $\pm$9.4&  286 $\pm$2.4 $\pm$14.3
           & 1.15 $\pm$0.01 $\pm$ 0.06& 272 $\pm$3 $\pm$27
           & 172 $\pm$3 $\pm$9\\
           & 2 & 67.8  $\pm$0.4 $\pm$6.8& 214 $\pm$2.4 $\pm$10.7
           & 1.19 $\pm$0.01 $\pm$ 0.06& 273 $\pm$2 $\pm$27
           & 178 $\pm$2 $\pm$9\\
           & 3 & 43.7  $\pm$0.3 $\pm$4.4 & 146.3 $\pm$2.2 $\pm$7.3
           & 1.25 $\pm$0.01 $\pm$0.06& 269 $\pm$3 $\pm$27
           & 172 $\pm$3 $\pm$9\\
           & 4 & 29.1  $\pm$0.2 $\pm$2.9& 101.1 $\pm$1.9 $\pm$10
           & 1.29 $\pm$0.01 $\pm$ 0.06&  263$\pm$3 $\pm$26
           & 168 $\pm$3 $\pm$9\\
    \hline
 K$^{+}$   & 0 & 20.1 $\pm$0.6 $\pm$1.9& 60.0 $\pm$4.7 $\pm$4.8& 1.00 
           $\pm$0.02 $\pm$0.05& 312 $\pm$13 $\pm$15 & 238 $\pm$7 $\pm$12\\
           & 1 & 16.2 $\pm$0.4 $\pm$1.5& 49.1 $\pm$4.3 $\pm$3.9 &
           1.05 $\pm$0.02 $\pm$0.05 & 307 $\pm$11 $\pm$15 &
           238 $\pm$6 $\pm$12\\
           & 2 & 10.9 $\pm$0.3 $\pm$1.0& 34.2 $\pm$3.5 $\pm$2.7&
           1.03 $\pm$0.01 $\pm$0.05& 309 $\pm$12 $\pm$15 &
           234 $\pm$6 $\pm$12\\
           & 3 & 6.9 $\pm$0.2 $\pm$0.7& 22.0 $\pm$2.8 $\pm$1.7 &
           1.06 $\pm$0.02 $\pm$0.05 & 291 $\pm$14 $\pm$15 &
           221 $\pm$7 $\pm$11\\
           & 4 & 4.2 $\pm$0.2 $\pm$0.4& 13.6 $\pm$2.3 $\pm$1.4&
           1.08 $\pm$0.03 $\pm$0.05 & 265 $\pm$16 $\pm$13 &
           208 $\pm$9 $\pm$11\\
    \hline
 K$^{-}$   & 0 & 8.5 $\pm$0.3$\pm$0.7& 21.0 $\pm$3.1 $\pm$2& 0.89
           $\pm$0.03 $\pm$0.05& 323 $\pm$19 $\pm$15 & 239 $\pm$10 $\pm$12\\
           & 1 & 6.2  $\pm$ 0.2$\pm$0.5& 15.9 $\pm$2.7 $\pm$1.3&
           0.91 $\pm$0.03 $\pm$0.05& 294 $\pm$16 $\pm$ 15 & 231 $\pm$9 $\pm$12\\
           & 2 & 4.7  $\pm$0.2$\pm$0.4& 11.4 $\pm$2.3 $\pm$0.9& 0.89
           $\pm$0.04 $\pm$0.05& 304  $\pm$16 $\pm$15 & 227 $\pm$8 $\pm$12\\
           & 3 & 2.7  $\pm$0.1 $\pm$0.2& 7.0 $\pm$1.8 $\pm$0.6& 0.88
           $\pm$0.03 $\pm$0.05&  275 $\pm$20 $\pm$14 & 207 $\pm$10 $\pm$11\\
           & 4 & 2.0  $\pm$0.1$\pm$0.2& 4.6 $\pm$1.5 $\pm$0.5& 0.85
           $\pm$0.04 $\pm$0.05& 291  $\pm$24 $\pm$15 & 219 $\pm$14 $\pm$11\\
  \end{tabular}
 \end{ruledtabular}
\vspace{-0.2cm}
 \caption
 {\label{PbPb-40} Summary of data for pion and kaon production in centrality
selected Pb+Pb collisions at 40\agev\ beam energy:
 rapidity density \dndy\ at mid-rapidity ($\pi^{-}$: $0<y<0.4$, K$^{\pm}$:
result from TOF analysis ($-0.1 < y < 0.1$)), total multiplicity $\langle N
\rangle$, RMS width of the
 rapidity distribution $RMS_{y}$, average transverse mass at mid-rapidity
$\langle \mt\rangle-m_{0}$, inverse slope parameter of \mt~spectra at
 mid-rapidity
($\pi^{-}$: $0<y<0.2$, fitrange 0.2 GeV/$c^{2}<\mt-m_{0}<$ 0.7
 GeV/$c^{2}$; K$^{\pm}$: $-0.1<y<0.1$ (TOF analysis), fit range 0
$<\mt-m_{0}<$ 0.8 GeV/$c^{2}$). The first error is statistical, the second
 systematic.}
\end{table*}
\begin{table*}[ht]
 \begin{ruledtabular}
  \begin{tabular}{ c c c c c c c}
  & Collision & \dndy\ & $\langle N \rangle$ & $RMS_{y}$ & $\langle
\mt\rangle
  -m_{0}$  & $T$ \\
  &  &  &  &  &  MeV/c$^{2}$ & MeV \\
    \hline
 $\pi^{-}$ & C+C & 3.25 $\pm$0.06 & 10.0 $\pm$1.4 
           & 1.18 $\pm$0.04 $\pm$0.1 & 247 $\pm$11 $\pm$25 & 169 $\pm$8 $\pm$9\\
           & Si+Si & 11.40 $\pm$0.19 & 34.6 $\pm$2.7
           & 1.16 $\pm$0.03 $\pm$0.1 & 243 $\pm$10 $\pm$24
           & 162 $\pm$6 $\pm$9\\
    \hline
 K$^{+}$   & C+C & 0.46 $\pm$0.06 & 1.26 $\pm$0.78  & 1.10
           $\pm$0.11 $\pm$0.15& &  \\
           & Si+Si & 1.82 $\pm$0.30& 5.1 $\pm$1.2 & 1.01
           $\pm$0.15 $\pm$0.15&  & \\
    \hline
 K$^{-}$   & C+C & 0.19 $\pm$0.02 & 0.44 $\pm$0.4 & 0.80
           $\pm$0.08$\pm$0.12 & & \\
           & Si+Si &  0.87 $\pm$0.07 & 2.0 $\pm$1.0 &
           0.89 $\pm$0.09$\pm$0.12& & \\
  \end{tabular}
 \end{ruledtabular}
\vspace{-0.2cm}
 \caption
 {\label{CC-SiSi-40} Summary of data for pion and kaon production in C+C and
Si+Si collisions at 40\agev\~beam energy:
 rapidity density \dndy\ at mid-rapidity ($\pi^{-}$: $0<y<0.4$, K$^{\pm}$: $y =
0$ average of the three fit functions), their multiplicities $\langle N \rangle$, 
and the RMS widths $RMS_{y}$ of
their rapidity distributions. For pions also average transverse masses $\langle
 \mt\rangle-m_{0}$ and the inverse slope parameters $T$ of the \mt~spectra are
given ($0<y<0.2$, fitrange 0.2 $<\mt-m_{0}<$ 0.7 GeV/$c^{2}$). 
The first error is the statistical, the second systematic.
The systematic errors on the yields are 15\%.}
\end{table*}

\begin{acknowledgments}
Acknowledgements: This work was supported by the US Department of
Energy Grant DE-FG03-97ER41020/A000, the Bundesministerium fur
Bildung und Forschung (06F 137), Germany, the Virtual Institute VI-146 of
Helmholtz Gemeinschaft, Germany, the Polish Ministry of Science
and Higher Education (1 P03B 006 30, 1 P03B 127 30,
0297/B/H03/2007/33, N N202 078735, N N202 204638 ), the Hungarian Scientific
Research Foundation (T032648, T032293, T043514), the Hungarian
National Science Foundation, OTKA, (F034707), the Bulgarian
National Science Fund (Ph-09/05), the Croatian Ministry of
Science, Education and Sport (Project 098-0982887-2878) and
Stichting FOM, the Netherlands.
\end{acknowledgments}


\end{document}